\documentclass[11pt,a4paper]{article}
\pdfoutput=1
\usepackage{jheppub}
\usepackage{graphicx}

\def\hbar{\hspace{0pt}\raisebox{1pt}{$-$} \hspace{-7pt} h}

\def\5{\overline 5}

\newcommand{\ba}{\begin{eqnarray}}
\newcommand{\ea}{\end{eqnarray}}
\newcommand{\no}{\nonumber}
\newcommand{\be}{\begin{equation}}
\newcommand{\ee}{\end{equation}}
\newcommand{\bea}{\begin{eqnarray}}
\newcommand{\eea}{\end{eqnarray}}

\def\EW{electro-weak~}

\title{The Natural Composite Higgs
}
\date{April 16, 2013
}
\author{
Luca Vecchi}
\affiliation{Maryland Center for Fundamental Physics,\\ Department of Physics, University of Maryland\\
College Park, MD 20742, USA}
\emailAdd{vecchi@umd.edu}
\abstract{The discovery of the Higgs boson has put considerable pressure on theories that aim to solve the hierarchy problem. Scenarios in which the Higgs is a pseudo-NGB of some new strong dynamics must possess a number of non-generic features in order to pass the progressively stringent collider bounds and simultaneously meet our naturalness criteria. Among these features are the existence of light fermionic partners of the top quark and an efficient collective breaking of the Nambu-Goldstone symmetry. The top partners have to be not only parametrically lighter than the other composites, but also weakly coupled to them in order to suppress unwanted flavor-violating effects. A Natural pseudo-NGB Higgs model should also be able to fit the LHC Higgs data without fine-tuning. Among theories with comparable compositeness scales, those that predict smaller corrections in the Higgs couplings to the standard model particles are therefore preferred. 

A concrete implementation of these ingredients is discussed in a scenario based on the coset $SU(5)/SO(5)$. The fit to the current LHC Higgs data is significantly improved compared to the minimal scenarios, and a fully natural explanation of both the weak scale and the Higgs boson mass can be attained. An important role is played by an independent quartic Higgs coupling generated by UV-sensitive loops involving electroweak doublets mixing with the top partners. The collider signature of this framework is shown to be rather model-dependent; in particular, the exotic scalars can alter the phenomenology of the top partners at a qualitative level. 
}
\begin{document}
\maketitle
%

\section{A ``LHC Paradox"?}
\label{intro}

The LHC has discovered a new boson of mass around $126$ GeV~\cite{Higgs!}. The couplings to the standard model (SM) particles are in such good agreement with those of a fundamental Higgs boson~\cite{HiggsFits}\cite{Montull:2012ik}\cite{Giardino:2013bma} that it gets harder and harder to convince oneself that this particle is not {\emph{the}} long sought Higgs boson, namely the last component of the weak doublet responsible for breaking the electroweak symmetry. What is not clear, though, is whether the Higgs boson in fact represents the first indication of an unknown, yet to be discovered {\emph{natural}} theory, or the only remaining ingredient of an {\emph{unnatural}} Higgs sector.

The concept of naturalness~\cite{naturalness} has been for decades one of the main driving principles in the Particle Physics community. And we believe this is for good reasons. Our understanding of the entire field is based on the notion of \emph{effective theory}, and all experimental data indicate that our best effective field theory -- the Standard Model -- is remarkably accurate up to a very large UV cutoff. The fact that the highest mass scale of the effective field theory is the weak scale $v\approx245$ GeV, and therefore many orders of magnitude smaller than the putative cutoff, is at the origin of the hierarchy problem, the flavor problem, and all sorts of theoretical puzzles. The point is that there is no symmetry within the Standard Model that can explain this huge hierarchy of scales. We therefore think such a symmetry must be introduced.

The most motivated theories for the weak scale have always been Supersymmetry and TeV scale Compositeness. However, after LEP it became evident that either our favorite natural theory was not the simple, minimal theory we hoped, or that our beloved scenarios were not completely natural after all. Yet, despite an impressive collection of indirect evidence, the Higgs boson was not found at LEP or the Tevatron. As a result, the ``LEP paradox" was not enough to discourage the model-builder from pursuing a natural description of electroweak symmetry breaking.

With the discovery of the Higgs one important piece has been added to the puzzle. The Higgs potential may now be parametrized by
\ba\label{V}
V=\frac{\mu^2}{2}h^2+\frac{\lambda}{4}h^4
\ea
with $v^2=-\mu^2/\lambda\approx(245$ GeV$)^2$ and $m_h^2=2\lambda v^2\approx(126$ GeV$)^2$, implying
\ba
\mu^2\approx-(89~{\rm GeV})^2.~~~~~~~~~\lambda\approx0.13.
\ea
Our {\emph{natural}} theory for the weak scale must now be turned into a theory for the entire Higgs sector: a theory aiming at naturally explaining $v$ should be able to explain $m_h$ without tuning the parameters. Naturalness has thus become a more stringent requirement, and the measurement of the total amount of fine-tuning should accordingly include this new, a priori independent, ``dimension". Formally, this statement means we should generalize the naturalness criteria of~\cite{Barbieri:1987fn} into something like 
\ba\label{deftuning}
{\rm fine~tuning}\sim\Delta_{v}\times\Delta_{m_h}\sim\frac{\delta\mu^2}{\mu^2}\times\frac{\delta\lambda}{\lambda}.
\ea
Here $\mu^2, \lambda$ are the measured values as given above, while $\delta\lambda,\delta\mu^2$ represent the variation of these couplings as a function of the fundamental parameters of the theory. $\Delta_{v,m_h}$ are simultaneously small in a theory with no fine-tuning.

With this new data point to explain, once more we find that our minimal scenarios are in serious trouble. The minimal supersymmetric standard model requires some non-minimal ingredients to naturally boost the Higgs boson mass up to $\sim126$ GeV. At the same time, the spectacular missing energy signatures predicted by the MSSM are not seen, thus pointing towards non-standard physics for the SUSY partners. 

The same conclusions apply to models of TeV Compositeness. Assuming the newly discovered particle is truly the Higgs, the good old Technicolor dream is definitively gone, and the only realistic possibility left are Composite Higgs scenarios in which the Higgs is a pseudo Nambu-Goldstone boson (NGB) of some new strongly coupled sector~\cite{CH}. Even so, a mass of the order $\sim126$ GeV is not what one would expect from generic strong dynamics. LEP has shown that the Higgs compositeness scale must be well above a few TeV, so unless new particles below the TeV are present, $\delta\mu^2$ and $\delta\lambda$ tend to be too large and the Higgs boson is predicted to be too heavy.

Are we facing a new ``LHC paradox"? Is the LHC telling us the weak scale is ``tuned"?

As opposed to LEP and the Tevatron, the LHC will be able to probe the Higgs sector all the way up to the TeV scale, which represents the ultimate energy where a fully natural theory for the weak scale can hide. This machine has thus the potential to change radically our understanding of the hierarchy problem, and with it of the Standard Model itself. But it seems a bit premature to give up naturalness now. It is instead timely to critically review our approach to model-building.

After just a few years of run of the LHC it has become apparent that the naturalness criteria may be made compatible with the experimental data only if many other theoretical prejudices are abandoned. Our theory should therefore be natural, but need not necessarily meet far less objective criteria such as minimality, simplicity, elegance, etc. 

In this view of the current status of Particle Physics, the best tool for the theoretician is the effective field theory approach. No matter how complicated the unknown short distance physics turns out to be, effective field theories have the ability to control its impact as long as this is not of immediate relevance for the currently probed energies.

This logic brings us to consider ``Natural" (or ``effective") SUSY, a supersymmetric scenario in which only the scalar partner of the top and a few other particles needed to naturally accommodate both the weak scale {\emph{and}} the Higgs mass are kept below the cutoff. This is perhaps our best chance to preserve the SUSY solution of the hierarchy problem without fine-tuning nor spoiling the remarkable agreement between the standard model and data.

The analogous framework for TeV Compositeness is offered by pseudo-NGB Higgs scenarios with sub-TeV top partners. Several papers have recently appeared on this subject~\cite{Berger:2012ec}\cite{Matsedonskyi:2012ym}\cite{Panico:2012uw}\cite{DeSimone:2012fs}. The present work is an effort in this same direction:

\vspace{0.5cm}


In section~\ref{sec:naturalCH} we analyze the main challenges the {\emph{Natural pseudo-NGB Higgs}} has to face, and stress some important features it should possess. There we emphasize the robust nature of the constraints arising from LHC searches, which {\emph{directly}} apply to the effective field theory as opposed to the bounds from flavor violation that are more sensitive to the UV completion. 

A crucial ingredient in these theories is the presence of light top partners. We will argue that these states not only have to be much lighter than the compositeness scale, but they should also couple {\emph{weakly}} to the strong dynamics, unless a highly non-generic flavor structure is postulated. In section~\ref{sec:picture} we therefore develop a systematic effective field theory expansion for models with Partially Composite top partners in which the role of the expansion parameter is played by the ratio between the mass scale of the top partners and the Higgs compositeness scale. The resulting low energy theory reduces in some limit to a ``Little Higgs" scenario, but with important differences.

The current experimental data together with our renewed notion of naturalness lead to a considerable amount of pressure on the effective field theory. This is enough for us to wonder if fully natural pseudo-NGB Higgs scenarios are still there at all. The bulk of the paper is devoted to addressing this issue. To better achieve our goal we decided to focus on an explicit model. This approach has the merit of providing a better feeling of ``how realistic" the scenario under study actually is. For reasons that will become clearer later, we chose the coset space $SU(5)/SO(5)$. 

Many of the important features identified in section~\ref{sec:naturalCH} will be found in this model, including small deviations in the Higgs couplings to the SM (section~\ref{sec:basics}), a natural explanation of the weak scale and the Higgs boson mass (section~\ref{sec:Hpotential}), and a rich collider phenomenology (section~\ref{sec:collider}). Despite the presence of potentially large corrections to the electroweak $T$ parameter, the model is found to cope rather well with electroweak data (section~\ref{sec:EWPT}). After a critical analysis of the model, we present our conclusions in section~\ref{sec:conclusions}.

\section{The Natural pseudo-NGB Higgs}
\label{sec:naturalCH}

In this section we will discuss old and new challenges for the Natural pseudo-NGB scenario. We will first briefly comment on the status of the experimental constraints {\emph{before}} the LHC in section~\ref{sec:fv}. We will then have a first look at the implications of the Higgs discovery on the fermionic sector (section~\ref{sec:f}), the gauge sector (section~\ref{sec:gauge1}), and the Higgs quartic coupling (section~\ref{HQ}) of these models. The most relevant collider constraints are then outlined in section~\ref{sec:directbounds}.

\subsection{Old problems: EW Precision Tests and Flavor Violation}
\label{sec:fv}

Historically, the main hurdles for scenarios of TeV scale Compositeness have been flavor violation and the \EW (EW) precision parameters. 

In models where the Higgs is a pseudo-NGB it is possible to separate the weak scale from the compositeness scale $m_\rho$, and the large tree-level corrections to the electroweak $S$ parameter may be taken under control if $m_\rho$ is larger than a few TeV. We will present a more careful study in an explicit model based on the coset $SU(5)/SO(5)$ in section~\ref{sec:EWPT}.

For what concerns flavor violation, it is nowadays widely believed that the best way to introduce a coupling of the SM fermions $f_{\rm SM}=q,u,d,\ell,e$ to the strong Higgs sector is via mixing operators of the form~\footnote{We suppressed the flavor indices for simplicity, but it should be understood that the ${\cal O}$s should come in at least 3 families if all the SM fermions are to acquire a mass this way.}
\ba\label{mix}
\lambda^{\cal O}_f\overline{f_{\rm SM}}{\cal O},
\ea
with ${\cal O}$ an operator that couples to the Higgs sector. In this sense the SM fermions are introduced as ``partially composite states". The Partial Compositeness idea was first proposed in the context of strongly coupled theories in~\cite{Kaplan:1991dc}, and then realized via the gauge/gravity correspondence as fermion wave function localization in 5D warped backgrounds~\cite{Gherghetta:2000qt}\cite{Huber:2000ie}.

The SM fermion masses will depend not only on the mixing $\epsilon_f\propto\lambda^{\cal O}_f$ between the SM fermions and the one-particle states created by ${\cal O}$, but also on the strength $g_{\cal O}$ controlling the coupling between the Higgs sector and ${\cal O}$. For example, the Yukawas for the up-type quarks will be $y_u\sim g_{\cal O}\epsilon_q\epsilon_u$, where $\sim$ means we ignore numbers $O(1)$.

Analogously, flavor violation beyond the renormalizable level arises from multiple insertions of~(\ref{mix}). For example, $\Delta F=2$ quark operators in generic theories are expected to be of the form ($\sim y_f^2$ is just a short for $\sim g_{\cal O}^2\epsilon_{f1}\epsilon_{f2}\epsilon_{f3}\epsilon_{f4}$)
\ba\label{FV}
\frac{y_f^2}{m_\rho^2}(\overline{f_{\rm SM}} f_{\rm SM})^2.
\ea
It has been recently shown in~\cite{KerenZur:2012fr} that a mass scale of order $m_\rho\sim10$ TeV would be required in the case the strong dynamics maximally violates flavor and CP. A scale of a few TeV will definitely suffice if the Higgs sector satisfies some approximate flavor symmetries. For recent work in this direction see for example~\cite{Redi:2011zi}\cite{Barbieri:2012tu}. Flavor observables in the lepton sector are especially constraining, but can be made compatible with data by pushing the lepton compositeness scale to higher energies than that associated to the quarks~\cite{Vecchi:2012fv}.

\subsection{The Top Partners}
\label{sec:f}

We will now argue that one can avoid large UV-sensitive corrections to the Higgs mass in Natural pseudo-NGB models in which the top partners have a higher compositeness scale than the Higgs boson (section~\ref{sec:fermion}), and then show that these latter fields must couple weakly to the Higgs dynamics in order to suppress flavor violation (section~\ref{sec:pc}). We will systematically describe them as Partially Composite states in section~\ref{sec:picture}.

\subsubsection{Collective Breaking}
\label{sec:fermion}

In generic pseudo-NGB Higgs models one expects the Higgs mass is sensitive to the compositeness scale $m_\rho$. However, electroweak precision measurements as well as flavor data push $m_\rho$ in the multi-TeV range, resulting in an unacceptably large $\mu^2$. The way out is introducing light partners for the SM fermions (most crucially the top) and gauge fields, that cut-off the quadratically divergent diagrams at a scale parametrically lighter than $m_\rho$. We here focus on the fermion sector, while the gauge sector will be discussed later on.

The cancellation of the quadratic divergence in the Higgs mass is due to a collective breaking of the Nambu-Goldstone shift symmetry~\cite{ArkaniHamed:2001nc}\cite{ArkaniHamed:2002qy}. Because we expect the SM fermions to be states external to the strong dynamics, any coupling involving them will generically break the global symmetry $G$ of the Higgs sector, and thus generate non-derivative couplings for the NGBs $\Pi$. To realize a collective breaking of the Higgs symmetry there must exist a field basis in which the Higgs couples only (i.e. dominantly) to the top partners, $Q$, and the SM fermions couple to $Q$ but not directly to the Higgs sector. The first coupling is introduced respecting a subgroup of $G$ that acts non-linearly on the NGBs. These requirements together with Partial Compositeness essentially determine the structure of the Lagrangian (in this reference basis)
\ba\label{EFT}
\lambda_q\overline{q}Q_R+\lambda_u\overline{u}Q_L+m_Q\overline{Q_L}P(\Pi)Q_R+\dots
\ea
Here we only consider $q$ and $u$, the $SU(2)_L$ doublet and singlet quarks respectively, since they are associated to the largest Yukawa couplings in the theory. An analogous coupling $\lambda_d\overline{d}Q_L$ should be present to give a mass to the down type quarks.

Collective breaking is now manifest: both mass parameters $m_Q$ and either $\lambda_q$ or $\lambda_u$ must be turned on to break the Nambu-Goldstone symmetry, and the spurionic charges of these couplings force the 1-loop contribution to the NGB potential to scale as
\ba\label{1'}
\frac{N_c}{16\pi^2}\lambda_f^2m_Q^2~\widehat V(\Pi)\left[1+O\left(\frac{\lambda^2_f}{m^2_Q}\right)\right].
\ea
Up to a model-dependent logarithmic sensitivity, the potential is thus controlled by mass scales much smaller than $m_\rho$, and $\mu^2$ can be naturally light if $m_Q,\lambda_f\ll m_\rho$. 

The top quark Yukawa arises from~(\ref{EFT}) after mixing with the $Q$s, precisely as explained after~(\ref{mix}):
\ba\label{y}
y_t\sim\frac{m_Q}{f}\epsilon_q\epsilon_u,
\ea
with $\epsilon_{q,u}\sim\lambda_{q,u}/m_Q$ the mixing angle between the SM fermions and the top partners. Using~(\ref{1'}),(\ref{y}), and $\epsilon_{q,u}\leq1$ we then obtain a lower bound on the Higgs mass squared:
\ba\label{1}
\delta\mu^2\gtrsim\frac{N_cy_t^2}{8\pi^2}m_Q^2\sim(90~{\rm GeV})^2\left(\frac{m_Q}{500~{\rm GeV}}\right)^2.
\ea
(A more accurate estimate of the tuning will have to wait until an explicit scenario is discussed, see section~\ref{EWSBnaturally} and figure~\ref{naturalness}.)

The lower bound in~(\ref{1}) is reached in theories in which both $\lambda_q$ and $\lambda_q$ are needed to break the Nambu-Goldstone symmetry. Now the potential will scale as $m^2_Q\lambda_q^2\lambda_u^2$ and will hence be finite. This always occurs if $P(\Pi)$ is a unitary matrix: switching off $\lambda_q$ ($\lambda_u$), the non-derivative couplings of the Higgs can be completely removed from~(\ref{EFT}) by a field redefinition of solely $Q_R$ ($Q_L$).~\footnote{An alternative way to saturate the lower bound in~(\ref{1}) is to introduce either $q$ or $u$ as resonances of the strong Higgs sector. In that case the composite component alone will respect $G$ and will not generate $\delta\mu^2$, such that loops involving both chiralities are needed. This possibility has been recently discussed in~\cite{Panico:2012uw}\cite{DeSimone:2012fs}.}

The Lagrangian~(\ref{EFT}) was proposed in the Little Higgs scenario of~\cite{Katz:2003sn}. One difference between our approach and the one usually adopted in the Little Higgs literature is that there $f$ is taken somewhat arbitrarily of the order of one TeV, which from~(\ref{y}) results in $m_Q\gtrsim y_tf\gtrsim O($TeV$)$. Here $f$ is viewed as a free parameter. 

A more qualitative difference is that in this paper we will ask which dynamics can possibly end up with the EFT~(\ref{EFT}) (section~\ref{sec:picture}), and how accurate that description actually is. The Lagrangian~(\ref{EFT}) cannot follow from a coupling like~(\ref{mix}), with ${\cal O}$ a generic operator of the strong dynamics and $Q$ one of its interpolating resonances. If that was the case $f_{\rm SM}Q$ would be replaced by unsuppressed couplings such as $f_{\rm SM}QP_{\rm new}(\Pi)$ which would lead to the usual quadratically divergence in the Higgs mass (unless $P_{\rm new}$ is unitary, in which case we are back to Eq.~(\ref{EFT})). To obtain a $\delta\mu^2$ with no residual power-law sensitivity on the scale $m_\rho$, the previous operators must be suppressed. One can avoid their presence, while still realizing Partial Compositeness, by cooking up a model where the mixing between the $Q$s and the SM fermions arises at scales {\emph{above}} $m_\rho$, and such that this mixing is the main channel via which the strong dynamics and $q,u$ talk. In this picture the $Q$s must have a compositeness scale higher than the Higgs, somewhat like a partially composite heavy fourth generation. 

There exist alternative options. However, our discussion of flavor violation in the next subsection will also point towards scenarios where the top partners are {\emph{not}} generic accidentally light composite states of the Higgs dynamics, but rather states {\emph{weakly coupled}} to the strong sector. It is this latter observation combined with the request~(\ref{EFT}) that led us to elaborate on the picture presented in section~\ref{sec:picture}.

\subsubsection{Flavor Violation and Light Top Partners}
\label{sec:pc}

Because collective breaking requires the $Q$s to be covariant under (part of) $G$, we may view these particles as light fermionic ``resonances" of the strong Higgs dynamics. There are then two obvious ways to explain why $m_Q\ll m_\rho$: the $Q$s are anomalously light chiral resonances that couple with full strength to the strong sector, or they are states that couple {\emph{weakly}} to the Higgs sector.

\paragraph{Anomalously light top partners and non-trivial Flavor}

The first option implies that non-chiral operators involving the $Q$s and the heavy states of the strong dynamics are unsuppressed. The dots of~(\ref{EFT}) for instance will include operators like~\cite{KerenZur:2012fr}
\ba
\frac{(\overline{Q}\gamma^\mu Q)^2}{f^2}\to\epsilon_f^4\frac{(\overline{f_{\rm SM}}\gamma^\mu f_{\rm SM})^2}{f^2}\sim\frac{y_f^2}{m_Q^2}(\overline{f_{\rm SM}}\gamma^\mu f_{\rm SM})^2,
\ea
where in the first step we rotated away the mixing $\lambda_f\overline{f_{\rm SM}}Q$, and in the second we employed~(\ref{y}). For $m_Q\lesssim1$ TeV, it is hard to pass the constraining flavor data even after the introduction of flavor symmetries~\cite{Barbieri:2012tu}. 

We thus see that unless more assumptions are made on how flavor is realized in the UV completion, the flavor problem cannot be decoupled in this case. One way to alleviate the bounds is to assume that the strong dynamics is invariant under a $U(3)_Q$ family symmetry and CP, such that all quark flavor and CP violation is encoded in $\lambda_{q,u,d}$. If no more structure is postulated, however, the 4$Q$ operators would be flavor conserving, but would still result in dangerous flavor-violating 4$f_{\rm SM}$ operators. The reason is ultimately that the 6 invariants $\lambda_{q,u,d}\lambda_{q,u,d}^\dagger, \lambda_{q}\lambda_{u,d}^\dagger, \lambda_{u}\lambda_{d}^\dagger$ that control flavor violation in the EFT are not in general in one-to-one correspondence with the SM Yukawas, such that flavor and CP violation are non-minimal. 

A best case scenario would be one where the number of flavor and CP parameters be exactly the same as in the SM. This is equivalent to ask that the physics responsible for generating the hierarchical parameters $\lambda_{q,u,d}$ imposes a constraint, say $\lambda_q={\cal F}(\lambda_{u},\lambda_{d})$, that is left invariant by a $U(3)^3$ subgroup of the original $U(3)_Q\times U(3)_q\times U(3)_u\times U(3)_d$ flavor symmetry. Depending on how the $U(3)^3$ is embedded, one finds different correlations among flavor observables.

A scenario that has been studied in~\cite{Cacciapaglia:2007fw}\cite{Redi:2011zi} is one that realizes minimal flavor and CP violation. In that scenario $\lambda_q\propto1_{3\times3}$, which is invariant under $U(3)^3=U(3)_{Q+q}\times U(3)_u\times U(3)_d$, so the 6 invariants above are just the Yukawa matrices and their polynomials. The problem with this framework is that the mixing of the three $q$ generations are the same, so the large top mass forces large corrections to the well measured properties of $u_L,d_L$. Indeed, for $\lambda_q\propto1_{3\times3}$ one abandons the paradigm of partial compositeness, by which the light generations are light because they are more weakly coupled to the strong sector.

A choice that avoids flavor problems and realizes the partial compositeness idea may be obtained by imposing the alternative ``GUT-like" constraint $\lambda_q\propto\lambda_d$. This constraint is invariant under $U(3)^3=U(3)_{Q}\times U(3)_u\times U(3)_{q+d}$ and the above 6 matrices $\lambda\lambda^\dagger$ become functions of the SM Yukawas, though not simple polynomials, such that the overall number of physical parameters is again the same as in the SM. A feature of this model is that $\Delta F=2$ 4-fermion operators in the down sector will first arise at 2-loop order in the $\lambda$s because their tree and 1-loop  coefficients are controlled by $(\lambda_q\lambda_q^\dagger)_{ij\neq i}\propto(\lambda_d\lambda_d^\dagger)_{ij\neq i}\propto(\lambda_q\lambda_d^\dagger)_{ij\neq i}\propto(y_d)_{ij\neq i}=0$. $\Delta F=1$ effects mediated by dipoles in the down sector also vanish up to 1-loop since the leading order contribution is aligned with the Yukawa. CP violation in flavor-conserving electric dipoles for $d,s,b$ vanish at the one-loop level if the strong dynamics respects CP because the Wilson coefficient of the relevant operators go as $\left[\lambda_q(\lambda^\dagger\lambda)\lambda_d^\dagger\right]_{ii}=\left[(\lambda_q\lambda^\dagger)(\lambda_q\lambda^\dagger)^\dagger\right]_{ii}$, which is real.

\paragraph{Partially composite top partners and anarchic Flavor}

If the $Q$s are weakly coupled to the strong dynamics the situation obviously improves, since now any time the Higgs sector generates an operator involving the $Q$s there will be a small parameter in front of it. We may think of the $Q$s as partially composite states, and identify the small parameter with the amount $\epsilon_Q$ of compositeness. Then according to NDA one gets (see section~\ref{sec:picture})
\ba\label{mQmrho}
m_Q\sim\epsilon_Q^2m_\rho.
\ea
This way by taking $\epsilon_Q\ll1$ we can explain the lightness of these states and thus keep the renormalization of $\mu^2$ under control as discussed in section~\ref{sec:fermion}. Similarly, flavor violation will be parametrically suppressed compared to the previous case:
\ba\label{FV1}
\epsilon_Q^4\frac{(\overline{Q} Q)^2}{f^2}\to\epsilon_Q^4\epsilon_f^4\frac{(\overline{f_{\rm SM}} f_{\rm SM})^2}{f^2}\sim\frac{y_f^2}{m_\rho^2}(\overline{f_{\rm SM}} f_{\rm SM})^2.
\ea
Eq.~(\ref{FV1}) shows the very same parametric scaling found in~(\ref{FV}). This conclusion generalizes to all flavor-violating operators, which are as discussed in the general scenarios of section~\ref{sec:fv}. The mass of the top partners does not enter, and can safely be taken to be much smaller than $m_\rho$ consistently with flavor data. Loops within the EFT~(\ref{EFT}) are down by $(m_\rho/4\pi f)^2\lesssim1$ compared to~(\ref{FV1}).  

As it has been in section~\ref{sec:fermion}, in obtaining~(\ref{FV1}) it was crucial that the dominant interactions between the SM fermions and the Higgs dynamics is mediated by partially composite  $Q$s. This assumption will be discussed in section~\ref{sec:picture}.

Our discussion suggests that the strong dynamics should feature a highly non-generic flavor structure if the $Q$s are anomalously light resonances of the Higgs sector, while a much more robust flavor framework is attained in models with partially composite $Q$s. For this reason we will focus on these latter scenarios in the following.

\subsection{The Gauge Sector}
\label{sec:gauge1}

In the gauge sector we expect
\ba\label{2}
\delta\mu^2\sim\frac{9g^2}{64\pi^2}m_G^2,
\ea
with $m_G$ denoting the mass scale of the gauge boson partners. Requiring $\delta\mu^2\lesssim(100$ GeV$)^2$ we obtain $m_G\lesssim1.3$ TeV. The scale $m_G$ can therefore be larger than $m_Q$, in which case the gauge boson partners would not be of immediate relevance for the current LHC searches. Even if they were lighter, however, the LHC phenomenology below $\sim1$ TeV would still be dominated by the physics of the top partners because the gauge boson sector is expected to be color neutral. 

In the following we specialize to an effective field theory (EFT) for the $Q$s, where the partners of the gauge bosons have been integrated out. The approach we follow here may be seen as a leading order approximation in the weak gauge couplings: for $g=g'=0$ our EFT is truly valid up to energies $\sim m_\rho$. 

The main impact of the gauge partners shows up in the Higgs potential and EW precision measurements. We will therefore discuss the former effect later on. To estimate the impact of the gauge partner sector on EW observables it is useful to have at least a sensible idea of what the UV completion may look like, and in particular what kind of physics could lead to a spectrum satisfying $m_Q<m_G<m_\rho$. A possibility is to construct a Little Higgs model where the mass scale of the top partners and that of the gauge partners are controlled by different parameters, as in~\cite{Schmaltz:2010ac}. An alternative would be to invoke SUSY. Suppose the Higgs sector is approximately supersymmetric, and that SUSY is softly broken in such a way that the lightest super-partners are a pseudo-NG higgsino, the wino and bino. In this framework it is still the $Q$s who cancel the quadratic divergence from the top sector, and the stop can safely be taken to be above the TeV scale along with the other scalars. $m_G$ will then coincide with the mass scale of the gauginos and higgsino(s). The model-building challenges here include avoiding too large tree-level corrections to the Higgs potential.~\footnote{The idea of combining Supersymmetry and a pseudo-NGB Higgs is sometimes referred to as ``double protection" in the literature~\cite{Birkedal:2004xi}\cite{Chankowski:2004mq}\cite{Roy:2005hg}\cite{Csaki:2005fc}. In these works the low energy physics is supersymmetric, and compositeness at $f\gtrsim O($TeV$)$ is used to cutoff a potentially large RG log from the cutoff down to the soft breaking mass scale. Our approach is different in that the hierarchy is entirely addressed by fermions at and below the TeV scale.}

Note that a $Z_2$ will be approximately satisfied by the gauge boson partners in both scenarios (T-parity~\cite{Cheng:2004yc} in the first class of scenarios and R-parity in the second), implying the absence of tree-level corrections to the electroweak parameters at scales $\sim m_G$, and possibly the existence of a dark matter candidate.  

We conclude that the precision EW data, the physics currently relevant to the LHC searches, as well as the Higgs couplings to the SM fermions and gauge bosons are all properties accurately described by our effective field theory below $m_G$. 


\subsection{The Higgs quartic coupling}
\label{HQ}

An independent Higgs quartic must be there to explain $v<f$ without resorting to a fine tuning of the parameters. Technically, the goal is to find a $G$-breaking spurion that generates a quartic but does not correct the Higgs mass. This problem was successfully addressed in the Little Higgs~\cite{ArkaniHamed:2001nc}, where it was emphasized that the NGB shift symmetry can be used to forbid a one-loop squared mass and simultaneously allow higher order couplings. 

Generally, one expects that one-loop diagrams involving the SM fields and their partners result in
\ba\label{gaugelambda}
\lambda\sim\frac{g_{\rm SM}^2}{16\pi^2}\frac{m_{\rm NP}^2}{f^2},
\ea
with $g^2_{\rm SM}$ a SM coupling (such as $y_t^2N_c$ or $g^2$) and $m_{\rm NP}$ a new physics threshold (the top or gauge partner masses, for instance). The observed value $\lambda=0.13$ \emph{requires perturbative new physics below the cutoff}, meaning that $m_{\rm NP}\ll4\pi f$ is necessary to keep the Higgs light. In this sense, we should discard scenarios with a ``tree-level" sized quartic. This conclusion is completely general, and applies to all composite Higgs models. 

Furthermore, naturalness arguments suggest that $m_Q$ be below a few hundred GeV, which together with $f<m_Q/y_t\sim m_Q$ tell us that $f$ cannot be much larger than the weak scale $v\approx245$ GeV. Again, this is perfectly consistent with a 1-loop order quartic. What we are looking for is a parametric $O(1)$ separation between the mass squared and the quartic (in units of $f$). 

In the $SU(5)/SO(5)$ model discussed in this paper we will take a bottom-up approach and look for spurion structures that can lead to a non-vanishing $\lambda$ without correcting the Higgs mass. We will find that the required structures naturally arise from loops of fermionic weak doublets that mix with the top partners. Perhaps not surprisingly, this fortunate effect is nothing but an accidental realization of the collective breaking mechanism of~\cite{ArkaniHamed:2001nc}.

\subsection{Direct bounds on the EFT}
\label{sec:directbounds}

\subsubsection{Top Partner Searches}
\label{sec:directbounds1}

Searches for pair produced, colored heavy fermions are currently underway at ATLAS and CMS, extending the Tevatron bounds up to about $m_Q\gtrsim500-800$ GeV, depending on the charges and branching ratios of the exotic state. The strongest constraints apply to final states with top quarks and hard leptons. For example, CMS finds a bound of $770$ GeV for BR$(Q\to W^+t)=100\%$ at the 8 TeV LHC with $\sim20/$fb of data by requiring a same sign dilepton pair and large transverse energy~\cite{CMS}. ATLAS, with a lower luminosity $\sim14/$fb, obtains a $95\%$ CL limit of $m_Q\gtrsim790$ GeV for a fourth generation quark doublet decaying into $th,Wb$ final states by simply selecting events with a hard lepton and many energetic jets (of which at least 2 $b$-jets)~\cite{ATLAS}.



The presently available analysis are already probing the region suggested by the naturalness criteria. The most sensible thing to do is therefore to focus on models in which the collective breaking of the Nambu-Goldstone symmetry is as effective as possible. In practice, this suggests looking among those models in which the bound in Eq.~(\ref{1}) is saturated, where the mass scale of the top partners may be pushed above the existing bounds without abandoning our main motivation for looking at the TeV scale. A complementary, not fully explored direction is to {\emph{relax}} the current bounds by constructing scenarios with a non-standard collider phenomenology for the $Q$s, in some analogy with what is done in Natural SUSY with R-parity violation.

\subsubsection{Higgs couplings at the LHC}
\label{sec:directbounds2}

For what concerns the Higgs boson properties, many authors have recently presented fits using the LHC Higgs data~\cite{HiggsFits}\cite{Montull:2012ik}\cite{Giardino:2013bma}. We employ the results of the most recent ref.~\cite{Giardino:2013bma}, that includes the latest news on $h\to\gamma\gamma$ from Moriond 2013 and also presents very useful simplified expressions for the $\chi^2$. 

\begin{figure}
\begin{center}
\includegraphics[width=2.7in]{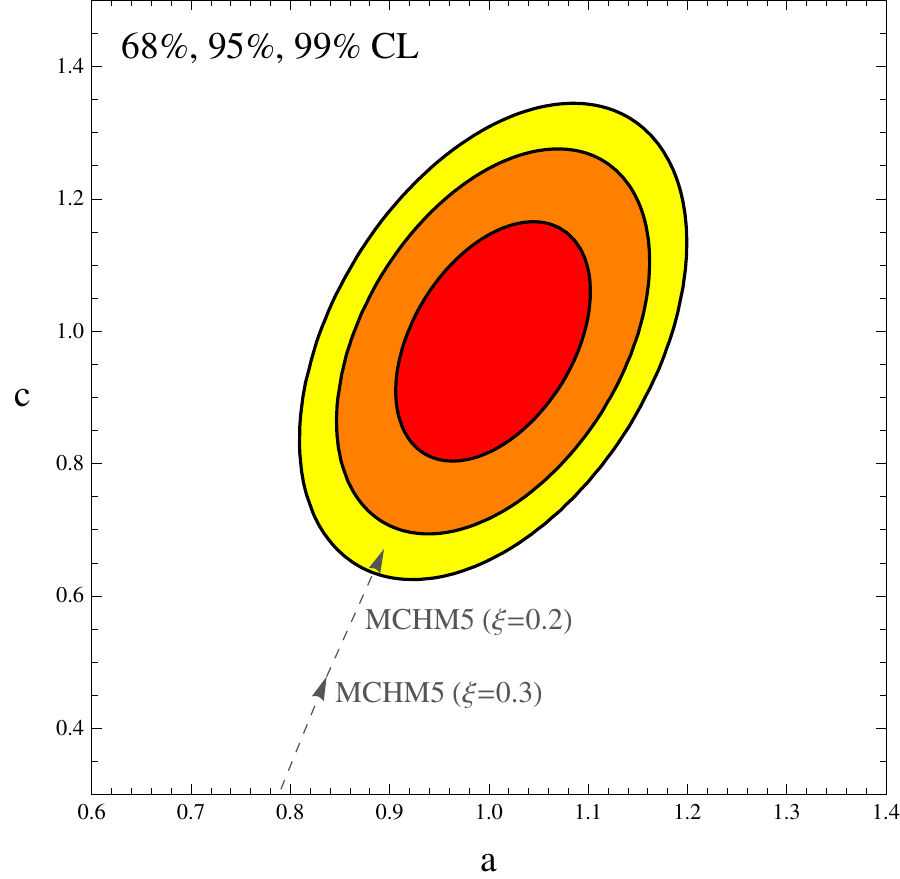}~~~~~~~~~~~\includegraphics[width=2.7in]{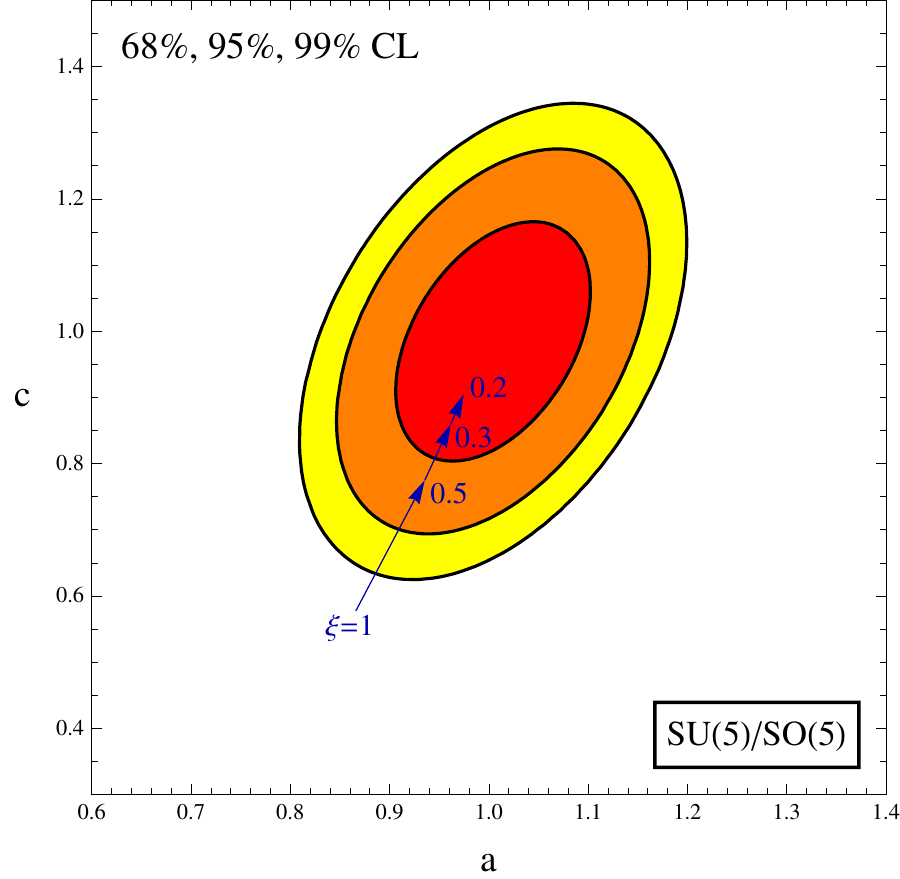}
\caption{\small $68\%, 95\%, 99\%$ CL contours for the Higgs couplings as defined in Eq.~(\ref{ac}). We use the ``universal" fit of~\cite{Giardino:2013bma} for 2 dof. On the left figure we superimpose the predictions of the minimal $SO(5)/SO(4)$ model with fermions embedded in the fundamental of $SO(5)$ for $\xi\equiv v^2/f^2=0.2,0.3$. On the right the same as left, but now for the $SU(5)/SO(5)$ model with $\xi=0.2, 0.3, 0.5, 1$. To make a sensible comparison, we defined the new physics scale $f$ in both models via the approximate relation $v\approx f\sin(\langle h\rangle/f)$, with $h$ the canonically normalized Higgs boson and $v=245$ GeV (see also footnote~\ref{foot:comp}).
\label{fitMCHM}}
\end{center}
\end{figure}

Parametrizing the fractional deviation in the coupling $hVV$ ($V=W^\pm,Z^0$) and $h\overline{f}f$ with two independent universal factors, 
\ba\label{ac}
a=\frac{g_{hVV}}{g_{hVV}^{\rm SM}}~~~~~~~~~~~~~c=\frac{g_{hff}}{g_{hff}^{\rm SM}},
\ea
we find the contours shown in figure~\ref{fitMCHM}. We see that the data are very well consistent with the couplings of a fundamental Higgs ($a=c=1$). The uncertainty is currently of order $20\%$ for vector bosons, and roughly twice as much for the coupling to fermions, which is mainly probed indirectly.

As seen in section~\ref{HQ}, a natural pseudo-NGB Higgs theory is characterized by a ``decay constant" $f$ just a factor of a few larger than the weak scale. This potentially leads to large corrections to the Higgs couplings $a=1-\xi c_H/2+O(\xi^2)$ and $c=1-\xi(c_H/2+c_y)+O(\xi^2)$~\cite{Giudice:2007fh}, with $c_{H}$ a numerical coefficient of order unity that depends on the algebra and $c_y=O(1)$ on the representation of the $Q$s. As a reference, the prediction of the minimal $SO(5)/SO(4)$ model with fermions embedded in a $5\in SO(5)$~\cite{Contino:2006qr} is shown for $\xi=v^2/f^2=0.2,0.3$ in figure~\ref{fitMCHM} on the left. The minimal scenario requires $f\gtrsim630$ GeV in order to pass the LHC constraints at the $95\%$ CL, and this translates into a bound $m_Q\gtrsim600$ GeV. The tension with the naturalness requirement gets more acute when the EW data are included. 

Models that for the same $f$ have numerically smaller $c_{H,y}$ have an advantage on the start. An example is the coset $SU(5)/SO(5)$ with $Q$s in the fundamental of $SU(5)$ that has $c_H=1/4$ and $c_y=3/8$ (see figure~\ref{fitMCHM} on the right), to be compared to the minimal $SO(5)/SO(4)$ scenario that has (for the same definition of $f$) $c_H=c_y=1$.

\subsection{Lessons for the Natural pseudo-NGB Higgs}
\label{sec:lessons}

\begin{figure}
\begin{center}
\includegraphics[width=2.8in]{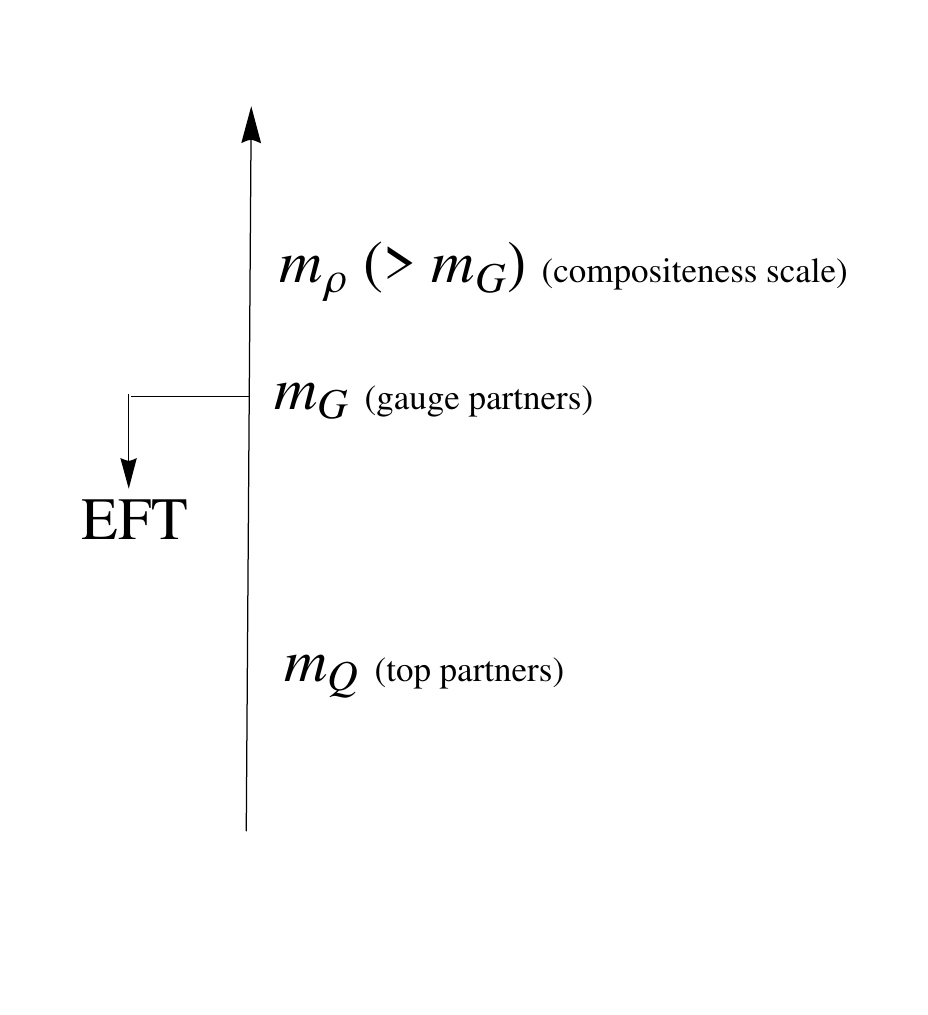}
\caption{\small Pictorial representation of the mass hierarchies in the Natural pseudo-NGB Higgs scenario. The highest mass is the compositeness scale, $m_\rho$, which sets the typical mass of the heavy composites and at which the NGBs are generated by the pattern $G\to H$. The gauge boson partners have masses $m_G<m_\rho$. Below this scale the theory reduces to an effective field theory for the SM and the top partners. Note that at leading order in the weak couplings $g,g'$ our EFT is truly valid all the way to $\sim m_\rho$.
\label{plot}}
\end{center}
\end{figure}

We have argued that below a scale of order $m_G\sim1.3$ TeV, a natural pseudo-NGB Higgs theory can effectively be described by a field theory for the SM, the composite Higgs, and the top partners $Q$ (see figure~\ref{plot}). Flavor violation is not a concern immediately applicable to the EFT. While the impact of EW precision observables should not be underestimated, the most robust {\emph{direct}} probes of this picture come from direct searches of the top partners as well as precision measurements of the Higgs properties. 

From the discussion in the previous subsections we conclude the Natural pseudo-NGB Higgs scenario must possess the following features:
\begin{itemize}
\item[i)] {\bf light, partially composite top partners} and\\ 
{\bf an efficient collective breaking} (section~\ref{sec:f})
\item[ii)] {\bf an independent Higgs quartic} (section~\ref{HQ})
\item[iii)] {\bf a versatile collider phenomenology} (section~\ref{sec:directbounds1})
\item[iv)] {\bf small corrections in the Higgs couplings to the SM} (section~\ref{sec:directbounds2})
\end{itemize}

Basically all of the above features point towards scenarios with non-minimal symmetry breaking patters. In particular, the collider phenomenology of the $Q$s may be qualitatively affected by decays into exotic NGBs, thus loosening the LHC constraints. This, together with the nice fit shown on the right of figure~\ref{fitMCHM}, motivated our study of the coset $SU(5)/SO(5)$, which we present starting in section~\ref{sec:basics}. Yet, before turning to this study we should develop an effective field theory approach for theories with partially composite $Q$s. This is done in the next section.

\section{EFT for Partially Composite Top Partners}
\label{sec:picture}

In section~\ref{sec:naturalCH} we emphasized the importance of having an EFT as in Eq.~(\ref{EFT}), with $Q$s weakly coupled to the Higgs sector and SM fermions interacting with the strong dynamics dominantly via the $Q$s. The aim of this section is to discuss one way in which these ingredients may be combined into a consistent framework. This exercise will provide us with important information about the expansion parameters characterizing these models.

We will focus on the physics of the SM fermions and the top partners. The gauge boson partners may be introduced along analogous lines.

The resulting framework will be somewhat in between conventional pseudo-NGB Higgs and Little Higgs scenarios. It is not an ordinary pseudo-NGB Higgs model because there exist fermionic partners of the top that are parametrically lighter than the compositeness scale and weakly coupled to the strong dynamics. It is not the typical Little Higgs scenario studied in the literature either, where a Lagrangian analogous to~(\ref{EFT}) is presented as an assumption. Since the $Q$s as well as the SM fermions must both couple to the strong sector somehow, it is not a priori clear under which assumptions the low energy Lagrangian will reduce to Eq.~(\ref{EFT}). We will see that in general Eq.~(\ref{EFT}) should in fact be supplemented with a UV-sensitive Higgs potential. It is this term that will bring a quartic Higgs coupling to the $SU(5)/SO(5)$ model (see section~\ref{sec:Hpotential}).

\subsection{The picture}

At very short distances the SM fermions couple to the strong dynamics via
\ba
\lambda_f^{\cal O}\overline{f_{\rm SM}}{\cal O},
\ea
with $\lambda_f^{\cal O}=\lambda_{q,u}^{\cal O}$ for $f_{\rm SM}=q,u,d,\ell,e$. At a lower energy $\sim\Lambda_Q$ we assume the strong dynamics can be described in terms of massless resonances $Q$ with appropriate charges to mix with the SM fermions, as well as other operators ${\cal Q}$. The most general Lagrangian renormalized at $\Lambda_Q$ will contain
\ba
{\cal L}_{\Lambda_Q}={\cal L}_{\rm kin}+\lambda_f\overline{f_{\rm SM}}Q+\lambda'_f\overline{f_{\rm SM}}{\cal Q}+\lambda_Q\overline{Q}{\cal Q}+\dots,
\ea
where ${\cal L}_{\rm kin}$ stand for the standard kinetic terms for $f_{\rm SM},Q$. The (dimensional) couplings $\lambda_f, \lambda_f'$ are expected to be of comparable strength, since both are controlled by $\sim\lambda_f^{\cal O}\Lambda_Q$, whereas $\lambda_f'/\lambda_Q$ is an independent parameter we will take to be $\ll1$. 

At this stage the global symmetry $G$ of the Higgs sector is linearly realized. The top partners $Q$ will thus have definite charges under it, which is already an important ingredient of~(\ref{EFT}). $G$ spontaneously breaks to $H\subset G$ at the scale $m_\rho\ll\Lambda_Q$, delivering the NGBs $\Pi$ with ``decay constant" $f$. The symmetry $G$ is however only approximate, because explicit broken by $\lambda_f^{\cal O}$ and the weak gauging of the SM subgroup (other sources of explicit breaking might be present as well). The NGBs will therefore acquire a potential controlled by these couplings.

A key observation is that for 
\ba\label{ass}
\frac{\lambda'_f}{\lambda_Q}\ll{\rm min}\left(\frac{\lambda_f}{m_\rho},1\right),
\ea
the dominant interaction between the Higgs sector and the SM fermions will be via the exchange of the $Q$s rather than the direct coupling $\lambda_f'$. The hierarchy~(\ref{ass}) was implicit in sections~\ref{sec:fermion} and~\ref{sec:pc}, and will be assumed in the following. Our analysis will be greatly simplified by this assumption.

At zeroth order in $\lambda_f'$, the effective Lagrangian obtained by integrating out {\emph{all}} fluctuations with momenta above $m_\rho$ reads:
\ba\label{EFTmrho}
{\cal L}_{\rm EFT}&=&{\cal L}_{\rm kin}+\lambda_f\overline{f_{\rm SM}}Q\\\no
&+&f^2m_\rho^2~\widehat{\cal L}_{\rm EFT}\left(\frac{\epsilon_Q Q}{f\sqrt{m_\rho}},\frac{\widehat\epsilon_f\epsilon_Qf_{\rm SM}}{f\sqrt{m_\rho}},\Pi,\frac{D_\mu}{m_\rho},\epsilon_Q,\widehat\epsilon_f\epsilon_Q,\cdots\right),
\ea
with ${\widehat{\cal L}}_{\rm EFT}$ an unknown function with $O(1)$ coefficients according to NDA, and $\cdots$ possible additional parameters. As usual (covariant) derivatives are suppressed by the relevant momentum, and $\Pi$ has zero weight.

The parameter $\epsilon_Q\leq1$ measures the amount of ``compositeness" of $Q$, that is how much $Q$ mixes with the Higgs sector at energies of order $m_\rho$. This coupling scales as $\sim\lambda_Q(m_\rho)/m_\rho$ for small $\lambda_Q(m_\rho)$ and suppresses all insertions of external $Q$ legs, but can also appear as an ordinary coupling constant (see the last entries in ${\widehat{\cal L}}_{\rm EFT}$) from diagrams where these fields are exchanged as virtual particles. 

Similar observations hold for the parameter $\widehat\epsilon_{f}\sim\lambda_{f}/m_\rho$, which represents the mixing between the SM fermions $f_{\rm SM}$ and the top partners at $m_\rho$. Note that under our assumption~(\ref{ass}), external $f_{\rm SM}$ legs must involve $Q$s, and hence come with a factor $\epsilon_Q\widehat\epsilon_f$.

As in the framework reviewed in section~\ref{sec:fv}, we are interested in models in which the composite field ${\cal Q}$ has a scaling dimension $d\gtrsim5/2$. This might be the result of a large anomalous dimension in a theory with no fundamental scalars, or simply because our theory has scalars $S$, in which case ${\cal Q}\sim\psi S$ for some fermion $\psi$. Under this generic condition the coupling $\lambda_Q$ is irrelevant, and one naturally expects $\epsilon_Q\ll1$ for $m_\rho\ll\Lambda_Q$: the $Q$s are partially composite in the sense of section~\ref{sec:pc}. On the other hand, $\lambda_f$ receives only small renormalization effects, so $\lambda_f(m_\rho)\sim\lambda_f$. We also expect $\lambda_f'/\lambda_Q$ to be approximately RG invariant, such that Eq.~(\ref{ass}) is truly a condition on the UV dynamics.

\subsection{The $m_Q/m_\rho$ expansion}

The Lagrangian~(\ref{EFTmrho}) has in general two expansion parameters, $\epsilon_Q$ and $\widehat\epsilon_f$. The former is always smaller than unity in order to suppress the couplings of the $Q$s to the Higgs sector, in particular their masses, as well as flavor violation in the visible sector. The other parameter $\widehat\epsilon_f$ is always small for the light generations, but may not be completely negligible for the top (see below). To be on the safe side we decide to expand~(\ref{EFTmrho}) in powers of $\epsilon_Q$:
\ba
{\cal L}_{\rm EFT}&=&{\cal L}_{\rm EFT}^{(0)}+\epsilon_Q^2{\cal L}_{\rm EFT}^{(1)}+\epsilon_Q^4{\cal L}_{\rm EFT}^{(2)}+O(\epsilon_Q^6).
\ea
We will now discuss the operators appearing in the leading orders of this expansion. Our analysis will not be exhaustive, but it will suffice to get some insight into the power counting underlying our analysis. 

First, note that ${\cal L}_{\rm EFT}^{(n)}$ will in general contain corrections to ${\cal L}_{\rm EFT}^{(n-1)}$. However, because $\epsilon_Q$ is singlet under $G$, higher powers of $\epsilon_Q$ will not induce new couplings for the NGBs. Because these are the main focus of the paper, we conclude we just need to worry about the leading order at which a given operator first appears.

At $O(\epsilon_Q^0)$ the Lagrangian simply reads
\ba\label{EFT0}
{\cal L}_{\rm EFT}^{(0)}={\cal L}_{\rm kin}+\lambda_f\overline{f_{\rm SM}}Q+\dots,
\ea
where the dots stand for operators involving the NGBs and the gauge fields. These have been discussed by a number of authors, see for instance~\cite{Giudice:2007fh}. There is no contribution to the Higgs potential because this is generated by insertions of $\lambda_f$, which always come with a $\epsilon_Q$ suppression.

At order $\epsilon_Q^2$, using our definition~(\ref{mQmrho}) $m_Q\sim\epsilon_Q^2m_\rho\ll m_\rho$, we find 2-fermion operators:
\ba\label{EFT2}
\epsilon_Q^2{\cal L}_{\rm EFT}^{(1)}&=&m_Q\overline{Q}P(\Pi)Q\\\no
&+&m_Q{\widehat\epsilon_f}^{~2}\overline{f_{\rm SM}}P_y(\Pi)f_{\rm SM}\\\no
&+&m_Q\widehat\epsilon_f\overline{f_{\rm SM}}P_Q(\Pi)Q\\\no
&+&\cdots,
\ea
plus analogous terms with derivatives, dipole operators, etc. In~(\ref{EFT2}) we focused on the parametric dependence of the couplings, and ignored factors of order unity.

In the first line of~(\ref{EFT2}) we recognize the operator which was used in~(\ref{EFT}) to realize collective breaking. They are accompanied by similar operators with derivatives of the NGBs, but these are typically subleading compared to those obtained at $O(\epsilon_Q^0)$ by removing the NGBs from~(\ref{EFT}) with a unitary, $\Pi$-dependent rotation of the $Q$s.

The most relevant operators arising from the class in the second line of~(\ref{EFT2}) are a Yukawa coupling for the SM fields (say up quarks):
\ba\label{yUV}
y\sim\frac{m_Q}{f}\widehat\epsilon_q\widehat\epsilon_u,
\ea 
and (possible) corrections to $\delta g_b$. There is another contribution to the Yukawa obtained by integrating out the $Q$s using the first line in~(\ref{EFT2}) and the mixing in~(\ref{EFT0}). This latter effect parametrically scales as
\ba\label{yIR}
y\sim\frac{m_Q}{f}\epsilon_q\epsilon_u,~~~~~~~~~~\epsilon_f\sim\frac{\lambda_f}{\sqrt{\lambda_f^2+m_Q^2}},
\ea
where we introduced the mixing $\epsilon_f$ between the SM fermions and $Q$ at scales $\sim m_Q$. Our claim is that~(\ref{yIR}) is larger, such that~(\ref{yUV}) may be ignored in a first approximation.

To prove this, note that the smallness of the light SM fermions translates into $\lambda_f\ll m_Q\ll m_\rho$, which says that~(\ref{yUV}) is in fact down by a factor $O(m_Q^2/m_\rho^2)$ compared to~(\ref{yIR}). For the top quark, and recalling that $m_Q>y_tf$, the generic expectation is that the coupling of at least one of the two chiralities is smaller than $m_Q$, say $\lambda_u<m_Q$. Then one finds~(\ref{yUV}) is parametrically suppressed compared to~(\ref{yIR}) by $m_Q/m_\rho$. 

Similar considerations apply to derivative operators in the second line of Eq.~(\ref{EFT2}), that can be shown to be negligible compared to the couplings obtained from the truncated Lagrangian~(\ref{EFT}), as well as the third line, which just induces a small correction to $\lambda_f\overline{f_{\rm SM}}Q$.

One can look at other operators and higher orders in $\epsilon_Q$, but the conclusion remains the same, up to some model-dependence: the effects proportional to $\widehat\epsilon_f$ typically represent small corrections to coefficients $\propto\epsilon_f$. For this reason our EFT at low energies approximately reduces to~(\ref{EFT}). 

There is an important exception to this rule, which is realized when the leading order term is anomalously small. In this case the corrections $\propto\widehat\epsilon_f$ might in fact be relevant. We find only one importance instance where this generically happens in the models of interest here: the Higgs potential.

In models where collective breaking is maximally efficient the Higgs potential will first arise at $O(\epsilon_Q^4)$. From~(\ref{EFTmrho}) we find $\epsilon_Q^4{\cal L}_{\rm EFT}^{(2)}\supset \delta V_{\rm EFT}$, with:
\ba\label{VEFT}
\delta V_{\rm EFT}=\frac{N_c}{16\pi^2}m_\rho^2m_Q^2(\widehat\epsilon_f\widehat\epsilon_{f'})^2~\widehat V(\Pi).
\ea
In order to determine whether this potential can be neglected we should compare it with the potential $O(\widehat\epsilon_f^{~0})$ generated by loops of $f_{\rm SM},Q$ with virtualities below $m_\rho$. In models where the bound in~(\ref{1}) is saturated, the latter effect is {\emph{finite}} and scales as $\frac{N_c}{16\pi^2}m_Q^4(\epsilon_q\epsilon_u)^2$, such that~(\ref{VEFT}) leads to a ``correction" of order
\ba\label{ratio}
\left(\frac{\widehat\epsilon_f}{\epsilon_q}\right)^2\left(\frac{\widehat\epsilon_{f'}}{\epsilon_u}\right)^2\frac{m_\rho^2}{m_Q^2}.
\ea
This term is potentially relevant when the third generation runs in the loop. Following the discussion below Eq.~(\ref{yIR}), and focusing for definiteness on the case $\lambda_u<m_Q$, we see that~(\ref{ratio}) will be much smaller than one for $\widehat\epsilon_f=\widehat\epsilon_{f'}=\widehat\epsilon_u$, and less or at most of order one for $\widehat\epsilon_f=\widehat\epsilon_q$ and $\widehat\epsilon_{f'}=\widehat\epsilon_u$. However, when $\widehat\epsilon_f=\widehat\epsilon_{f'}=\widehat\epsilon_q$ the effect may be as large as $(\epsilon_q/\epsilon_u)^2(m_\rho/m_Q)^2$, and can easily dominate!

We conclude that the EFT in~(\ref{EFT}) provides an accurate description of theories with partially composite $Q$s if supplemented with the potential term $\delta V_{\rm EFT}$ in~(\ref{VEFT}).

\section{The $SU(5)/SO(5)$ pseudo-NGB Higgs}
\label{sec:basics}

The coset $SU(5)/SO(5)$ was proposed in~\cite{Georgi:1984af} as one of the first examples of pseudo-NGB Higgs theories with custodial symmetry. The model was then reconsidered in the context of the ``Little Higgs" in~\cite{ArkaniHamed:2002qy}. There the top partners were introduced, but the custodial symmetry was explicitly broken in order to realize a collective breaking mechanism in the gauge sector. The effective field theory we will study in this paper has first appeared in the ``Intermediate Higgs" of~\cite{Katz:2005au}. The custodial symmetry is restored, and the partners of the gauge fields are decoupled. 

Our analysis differs from that of~\cite{Katz:2005au} in several respects. First, we do not identify the scale of the gauge boson partners $m_G$ with the compositeness scale $m_\rho$, as stressed in section~\ref{sec:gauge1}, and this inevitably requires new structure above $m_G\sim1.3$ TeV. Second, our study will include the recent LHC constraints on the Higgs couplings and the top partners, as well as an analysis of the electroweak precision data and the physics of the EW triplet NGB. Third, we will study the various contributions to the Higgs potential in much more detail than done in~\cite{Katz:2005au}, where several other symmetry breaking patterns where discussed.

\subsection{Basics on $SU(5)/SO(5)$}

The symmetry breaking pattern $SU(5)\to SO(5)$ may be thought of as arising from the vacuum of a two-index symmetric representation of $SU(5)$. The adjoint is easily seen to split into the unbroken generators $T^\alpha$ of $SO(5)$ and the broken ones $X^a$, which reside in a symmetric traceless $14\in SO(5)$. This delivers $14$ NGBs transforming as 
\ba\label{NGB1}
(3,3)\oplus(2,2)\oplus(1,1).
\ea
under $SU(2)_L\times SU(2)_R\subset SO(5)$. These further decompose under the SM $SU(2)_L\times U(1)_Y$ gauge group as ${3}_{\pm1}+{3}_0+{2}_{\pm1/2}+{1}_0$.

Following the standard technique~\cite{CCWZ} we describe the NGBs $\Pi=\Pi^a X^a$ with a unitary matrix $\xi(\Pi)$ transforming {\emph{non-linearly}} under $SU(5)$ as $\xi(\Pi)\to g\xi(\Pi) h^\dagger(\Pi,g)$, with $g\in SU(5)$ and $h\in SO(5)$. The algebra of the generators satisfies relations of the form 
\ba
[T,T]\propto T,~~~~~~~[T,X]\propto X,~~~~~~~[X,X]\propto T. 
\ea
The first simply represents the closure of the $SO(5)$ algebra, the second defines the transformation properties of the NGBs, while the last relation says that the coset is symmetric. This latter fact ensures the presence of two automorphisms~\footnote{We refer the reader to ref.~\cite{Gripaios:2009pe}, of which we also borrow the notation, where a similar analysis was presented for the coset $SU(4)/Sp(4)$.} $A_1: (T,X)\to (T,-X)$ and $A_2: (T,X)\to (-T^t, X^t)$. These two automorphisms have important implications. First, they entail symmetries of the effective field theory (up to possible anomalies). Second, they result in a simplification of the Maurier-Cartan one-form, which in practice means one can arrange the NGBs into a matrix $U=e^{i\Pi}$ transforming {\emph{linearly}} under $SU(5)$:
\ba
U\equiv\xi\tilde\xi^\dagger\to gU\tilde g^\dagger.
\ea
Here $\tilde\xi=\xi^\dagger$ is the image of $\xi$ under the automorphism $A_1$. The kinetic term of the NGBs then simply reads
\ba\label{NGBkinetic}
{\cal L}_{\rm NGB}=\frac{f^2}{4}{\rm tr}[D_\mu U D^\mu U^\dagger],
\ea
where $D_\mu$ is the electroweak covariant derivative, which can be straightforwardly written down by observing that $U\to hUh^\dagger$ for $h\in SO(5)$,  and $f$ is a mass parameter controlling the strength of the NGB interactions.

We choose to embed $SU(2)_L\times SU(2)_R\subset SU(5)$ by writing the defining $5\in SU(5)$ as
\ba\label{embed1}
\left( \begin{array}{c}  \psi_-  \\
\psi_+ \\ \psi_0 \end{array}\right).
\ea
Here $(\psi_-,\psi_+)$ transforms as a $(2,2)\in SU(2)_L\times SU(2)_R$: 
\ba\label{(2,2)}
\psi^i_a\equiv\left( \begin{array}{cc}  \psi_-^u~~\psi_+^u  \\
\psi_-^d~~\psi_+^d \end{array}\right)\to L_{ij}\psi^j_bR^\dagger_{ba}
\ea
while $\psi_0$ is a $(1,1)$. It is sometimes more convenient to think of the embedding~(\ref{embed1}) as the following identification 
\ba\label{embed2}
\left( \begin{array}{cc}  
L\otimes R^*&\quad \\
\quad&1
\end{array}\right)\in SU(2)_L\times SU(2)_R,
\ea
where the tensorial product $\otimes$ can immediately be read off from Eq.~(\ref{(2,2)}). We identify the three generators of $SU(2)_L$ and $T_R^3$ with the generators of the SM gauge group. Note the presence of a global $SU(2)_L\times SU(2)_R$ custodial symmetry (that was explicitly broken by gauge couplings in~\cite{ArkaniHamed:2002qy}), spontaneously broken down to its diagonal by the Higgs $H\sim(2,2)$.

The vacuum may be parametrized by a unitary and symmetric matrix (real by CP invariance)
\ba
\Sigma_0=\left(\begin{array}{ccc} 
\quad&-\epsilon&\quad \\
\epsilon&\quad&\quad\\
\quad&\quad&~~1 
\end{array}\right)~~~~~~~~~~~\Sigma_0^t=\Sigma_0~~~~~~~~~\Sigma_0^2={\bf{1}}_{5\times5},
\ea
with $\epsilon$ the 2 by 2 antisymmetric tensor of $SU(2)$. By construction $\Sigma_0$ is invariant under $SO(5)$, $T\Sigma_0+\Sigma_0T^t=0$. Analogously one gets $X\Sigma_0-\Sigma_0X^t=0$. These relations are encoded in the following useful expression:
\ba\label{useful}
g^*\Sigma_0=\Sigma_0 \tilde g.
\ea

We now have all the tools needed to write down an explicit expression for the NGB matrix. From~(\ref{NGB1}), and employing our choice of embedding~(\ref{embed1})(\ref{embed2}), we derive
\ba\label{NGB}
f\Pi=\left( \begin{array}{ccc}  
\phi_0+\frac{\eta}{\sqrt{10}}{\bf 1}_{2\times2}&-i\phi_+^\dagger&i\tilde H  \\
i\phi_+&-\phi_0+\frac{\eta}{\sqrt{10}}{\bf 1}_{2\times2}&iH\\
-i\tilde H^\dagger&-iH^\dagger&-4\frac{\eta}{\sqrt{10}} 
\end{array}\right)\\\no
\ea
Here $\phi_{+}=\sigma^a\phi_{+}^a$ is a complex (i.e. $(\phi_+^a)^*\neq\phi_+^a$) EW triplet with hypercharge $+1$ while $\phi_{0}=\sigma^a\phi_{0}^a$ is a real ($(\phi_0^a)^*=\phi_0^a$) EW triplet with hypercharge $0$; $H$ is the Higgs doublet of hypercharge $+1/2$ while $\tilde H=\epsilon H^*$. 

We defined CP as the action of both automorphisms $A_1\cdot A_2$,
\ba\label{A1}
H\to H^*~~~~~~~~~\phi_+\to\phi_+^*~~~~~~~~~\phi_0\to-\phi_0^*~~~~~~~~~\eta\to-\eta,
\ea
 together with space inversion. The powers of $i$ in~(\ref{NGB}) are chosen to ensure the Higgs boson is CP-even.

\subsection{Matter fields}

To ensure the realization of collective breaking, the top partners must be color triplets, and transform under some representation of the group $SU(5)\times U(1)_X$, where the abelian factor remains unbroken at $m_\rho$ and is used to accommodate the appropriate charges under hypercharge.

We introduce a left-handed and a right-handed fermionic resonance with charges $\sim5_{2/3}$ and $\sim\overline{5}_{2/3}$, respectively. Other representation might be considered, but we decided to focus on this minimal choice. We opt to describe the $Q$s in terms of an EW vector-like Dirac field with definite SM quantum numbers. To this end we introduce the following notation:
\ba
Q_L\equiv P_L Q~~~~~~~~~~~~~~~Q_R\equiv\Sigma_0 P_RQ~~~~~~~~~~~~
Q=\left( \begin{array}{c}  
Q_-\\
Q_+\\
Q_0\end{array}\right),
\ea
where $P_{L,R}$ are left or right chirality projectors and $Q$ is a Dirac fermion with components transforming \emph{non-homogeneously} under $SU(5)$. Specifically, the left handed and right handed components of $Q$ transform respectively as $Q_L\to gQ_L$ and $Q_R\to\tilde gQ_R$ under $SU(5)$. Using~(\ref{useful}) we see that, as anticipated, $Q_{L,R}\to hQ_{L,R}$ for $h\in SO(5)$. In other words, $Q_{+,-}$ are $SU(2)_L$ doublets of hypercharge $7/6,1/6$ and $Q_0$ is an $SU(2)_L$ singlet with hypercharge $2/3$.

In section~\ref{sec:picture} we found that the relevant Lagrangian at quadratic order in the fermion fields is:
\ba\label{EFT1}
{\cal L}_{\rm EFT}&=&{\cal L}_{\rm kin}+\lambda_q \overline{q}Q_R+\lambda_u\overline{u}Q_L+{\rm hc}\\\no
&+&\frac{f^2}{4}{\rm tr}[D_\mu U D^\mu U^\dagger]+m_Q\overline{Q}_LU Q_R-\delta V_{\rm EFT}\\\no
&+&\dots
\ea
Note that a mass term for the $Q$ (without insertions of $U$) is forbidden by $SU(5)$ invariance. Terms involving two $Q$s and (derivatives of) the NGBs turn out to be subleading (order $\epsilon_Q^2$) with respect to those already in~(\ref{EFT1}). Analogously, other interactions among the SM fermions, the top partners, and the NGBs are suppressed by powers of $\widehat\epsilon_{q,u}$, and will be neglected (see section~\ref{sec:picture} for more details). The Higgs potential $\delta V_{\rm EFT}$ will be carefully discussed in section~\ref{sec:Hpotential}.

To keep track of the $SU(5)$ indices it is convenient to think of $\lambda_{q,u}$ as ``spurions" with values $\lambda_q=\bar\lambda_q\Delta_q$ and $\lambda_q=\bar\lambda_q\Delta_q$, where $\bar\lambda_{q,u}$ are $SU(5)$-singlet matrices in flavor space while $\Delta_{-,0}$ are $SU(5)$-charged fields with the following backgrounds
\ba
\Delta_-=\left({\bf 1}_{2\times2}\quad{\bf 0}_{2\times2}\quad{\bf 0}_{2\times1}\right)~~~~~~~~~~~~~~\Delta_0=\left({\bf 0}_{1\times2}\quad{\bf 0}_{1\times2}\quad1\right).
\ea
These expressions select the $Q_{-,0}$ components of the $Q$s, respectively, and will turn out to be very useful when deriving the Higgs couplings. 

The automorphism $A_1$ acts as $Q_L\leftrightarrow Q_R$ (as usual up to phase transformations) and is violated by the couplings with the SM.

The EFT~(\ref{EFT1}) saturates the bound in Eq.~(\ref{1}) because $P(\Pi)=U$ is unitary. Let us now look at the couplings of the Higgs boson in some detail.

\subsection{Modified Higgs Couplings}
\label{sec:MHC}

The leading couplings of the NGBs to the SM can be described by an effective Lagrangian where the heavy fermions have been integrated out. The exact tree-level integration of the $Q$ is shown in Eq.~(\ref{LET}). Assuming for simplicity that $m_Q$ is approximately flavor universal, we canonically normalize the SM fields in~(\ref{LET}) and finally obtain:
\ba\label{EFTbelowQ}
{\cal L}^{\mu<m_Q}_{\rm EFT}&=&\frac{f^2}{4}{\rm tr}[D_\mu U D^\mu U^\dagger]-\delta V_{\rm EFT}\\\no
&+&\overline{q}\gamma^\mu iD_\mu q+\overline{u} i\gamma^\mu iD_\mu u\\\no
&-&\epsilon_q\overline{q}m_QU^\dagger u\epsilon_u^\dagger+{\rm hc}\\\no
&+&\epsilon_q\overline{q}\gamma^\mu (U^\dagger iD_\mu U) q\epsilon_q^\dagger+\epsilon_u\overline{u} \gamma^\mu (U iD_\mu U^\dagger)u\epsilon^\dagger_u\\\no
&+&\cdots,
\ea
where the remainder refers to $O(p^2)$ operators involving the SM fermions, and
\ba\label{epsilonq}
\epsilon_q=\frac{\lambda_q}{\sqrt{\lambda_q\lambda_q^\dagger+m_Q^2}}~~~~~~~~~~~~~~~~~~~\epsilon_u=\frac{\lambda_u}{\sqrt{\lambda_u\lambda_u^\dagger+m_Q^2}}
\ea
are the sine of the mixing angles between the $Q$s and the SM quarks $q,u$ respectively.

To find explicit expressions for the masses of the vector bosons and the SM fermions we set $\phi_0=\phi_+=\eta=0$ in~(\ref{NGB}) and work in the unitary gauge $H^t=(0,h/\sqrt{2})$. In this limit it is easy to derive:
\ba
U=\left( \begin{array}{ccccc}  
\frac{\cos+1}{2}~&0&~~0&~\frac{\cos-1}{2}&-\frac{\sin}{\sqrt{2}}\\
0~&1&~~0&~0&0\\
0~&0&~~1&~0&0\\
\frac{\cos-1}{2}~&0&~~0&~\frac{\cos+1}{2}&-\frac{\sin}{\sqrt{2}}\\
\frac{\sin}{\sqrt{2}}~&0&~~0&~\frac{\sin}{\sqrt{2}}&c
\end{array}\right)
\ea
with $\sin\equiv\sin(h/f)$ and $\cos\equiv\cos(h/f)$. Inserting this matrix back into~(\ref{NGBkinetic}) we get
\ba\label{mW1}
{\cal L}_{\rm NGB}=\frac{1}{2}(\partial_\mu h)^2+m_W^2(h)\left[W^+_\mu W^-_\mu+\frac{1}{2\cos\theta_w}Z^0_\mu Z^0_\mu\right],
\ea
with~\cite{Georgi:1984af}
\ba\label{mW}
m_W^2(h)=\frac{g^2}{2}f^2(1-\cos).
\ea
We see that in our notation the Higgs boson $h$ is canonically normalized, and the physical mass of the $W^\pm$ boson may be written as
\ba
m_W^2=\frac{g^2}{2}f^2(1-\langle \cos\rangle)\equiv \frac{g^2}{4}v^2,
\ea
where $v\approx245$ GeV.

As explained in~\cite{Giudice:2007fh}, the very fact that the SM fermion and vector boson masses in~(\ref{mW1}) are not linear in $h$ implies that the couplings of the composite Higgs boson will differ from those of a fundamental field. The deviations from the SM expectation are parametrized by the couplings $a,c$ defined in Eq.~(\ref{ac}), and parametrically scale as powers of $v^2/f^2$ times a model-dependent numerical coefficient.

From~(\ref{mW}) we derive
\ba\label{a}
a\equiv\frac{g_{hWW}}{g_{hWW}^{\rm SM}}=\left.\frac{1}{gm_W(h)}\frac{\partial m_W^2(h)}{\partial h}\right|_{h=\langle h\rangle}=\sqrt{1-\frac{\xi}{4}}.
\ea
where we adopted the definition 
\ba\label{xidef}
\xi\equiv\frac{v^2}{f^2}=2(1-\langle \cos\rangle).
\ea
The numerical coefficient in front of the $O(\xi)$ correction is small, basically four times smaller than the value $\sqrt{1-\xi}$ of the minimal cosets $SO(5)/SO(4)$~\cite{Agashe:2004rs} and $SU(4)/Sp(4)$~\cite{Galloway:2010bp}\cite{Gripaios:2009pe}.~\footnote{This statement deserves an explanation, since the normalization of $f$ is model-dependent. For example, there is a factor of $\sqrt{2}$ difference in our definition and the one adopted in~\cite{Katz:2005au}. A meaningful comparison between models can be done if the parameter $f$, seen as a function of $v,\langle h\rangle$, is defined the same way. In refs.~\cite{Agashe:2004rs}\cite{Galloway:2010bp}\cite{Gripaios:2009pe} the authors employed the definition $\xi_{\rm lit}=\langle \sin^2\rangle$. With our convention~(\ref{xidef}) we find
\ba
\langle \sin^2\rangle=\xi\left(1-\frac{\xi}{4}\right),
\ea
such that $\xi=\xi_{\rm lit}+\xi_{\rm lit}^2/4+\dots$. Up to small corrections the two definitions coincide.\label{foot:comp}} 
This has important implications that we will discuss shortly, and in section~\ref{sec:EWPT}.

We can now repeat the analysis for the fermion sector. Here it is important to stress that the fermion wave-functions arising from our tree-level integration of the $Q$s do not depend on the NGBs. From this seemingly meaningless observation follows that $c$ is {\emph{flavor universal}}.~\footnote{In section~\ref{sec:picture} we argued that there can be additional contributions to the Yukawa couplings of the third generation fermions coming from physics above $m_\rho$. These are expected to be of order $\sim m_Q/m_\rho$, but have no definite sign, so could result in an enhancement as well as a suppression of the top and bottom Yukawas. We decide to stick to the minimal, more predictive regime $m_Q\ll m_\rho$ in what follows.}

The SM fermion masses follow from~(\ref{EFTbelowQ}), and read
\ba\label{mf}
m_f(h)=\frac{{\epsilon_q}m_Q{\epsilon_u}^\dagger}{\sqrt{2}} \sin.
\ea
This gives
\ba\label{c}
c\equiv\frac{g_{hff}}{g_{hff}^{\rm SM}}=\left.\frac{2m_W(h)}{gm_f(h)}\frac{\partial m_f(h)}{\partial h}\right|_{h=\langle h\rangle}=\sqrt{\frac{1-\frac{\xi}{2}}{1+\frac{\xi}{2}}}\leq1.
\ea
This value is roughly three times smaller than in the minimal $SO(5)/SO(4)$ model with fermions in the fundamental of $SO(5)$~\cite{Contino:2006qr}, or $SU(4)/Sp(4)$ with matter in the antisymmetric $6\in SU(4)$~\cite{Gripaios:2009pe}.

The coupling to gluons receives corrections from physics above and below the cutoff. The first may be parametrized by adding $H^\dagger H G_{\mu\nu}G^{\mu\nu}$ to the EFT~(\ref{EFT1}). Because this operator breaks the Nambu-Goldstone symmetry, we estimate its coefficient to be of order $(g_s^2/m_\rho^2)\times\delta\mu^2/m_\rho^2$, in agreement with~\cite{Low:2009di}. The resulting effect is negligible. Loops of the top partners can be estimated analogously, and are naively expected to renormalize the SM coupling $\sim g_s^2/16\pi^2 v^2$ at order
\ba
\frac{\delta\mu^2}{m_Q^2}\frac{16\pi^2 v^2}{m_Q^2}\sim\frac{m_t^2}{m_Q^2}\ll1.
\ea
The dominant correction to the coupling $hgg$ comes in fact from the modified Higgs coupling to the top. Using the Higgs low energy theorems~\cite{Ellis:1975ap} we indeed find that it is just a factor $c$ smaller than the SM (see Eq.~(\ref{hgg})). This feature of pseudo-NGB Higgs models has been studied by several groups, including~\cite{Low:2009di}\cite{Low:2010mr}\cite{Azatov:2011qy}\cite{Gillioz:2012se}\cite{Delaunay:2013iia}.

Similarly, the dominant correction to $\Gamma(h\to\gamma\gamma)$ is due to $a,c$. In addition, there are potentially relevant loops of the new NGBs:\footnote{Ref.~\cite{Chala:2012af} considered this effect in the (non-symmetric) coset $SO(7)/G_2$.} 
\ba
\frac{\Gamma(h\to\gamma\gamma)}{\Gamma_{\rm SM}(h\to\gamma\gamma)}=
\left|\frac{A_1(\tau_W)a+A_{1/2}(\tau_t)c+\sum_\phi Q_\phi^2N_\phi\frac{\lambda_{h\phi\phi}v^2}{2m_\phi^2}A_0(\tau_\phi)}{A_1(\tau_W)+A_{1/2}(\tau_t)}\right|^2.
\ea
Here $\tau_i=4m_i^2/m_h^2$, while the loop functions are defined as (see for instance~\cite{Carena:2012xa}):
\ba
A_1(x)&=&-2-3x\left[1+(2-x)f(x)\right]\\\no
A_{1/2}(x)&=&\frac{4}{3}\times2x\left[1+(1-x)f(x)\right]\\\no
A_0(x)&=&-x\left[1-xf(x)\right],
\ea
where for $x>1$:
\ba
f(x)=\arcsin^2\left(1/\sqrt{x}\right).
\ea
The coupling $\lambda_{h\phi\phi}$ is defined by
\ba
V_{\rm total}\supset\frac{\lambda_{h\phi\phi}}{2}(\langle h\rangle+h)^2|\phi|^2,
\ea
where $V_{\rm total}$ is the effective scalar potential. $N_\phi$ and $Q_\phi$ are the number and electric charge of the scalars running in the loop. We find that terms $O(\Pi^4)$ in $V_{\rm total}$ do not contain the operator $H^\dagger\phi^\dagger\phi H=\tilde H^\dagger\phi\phi^\dagger\tilde H$ that would otherwise lead to trilinear couplings involving the doubly charged component of $\phi_+$. Hence, only the singly charged components of $\phi_{+,0}$ enter, and we have $Q_\phi=1$ and $N_\phi=2$. (A coupling between the doubly charged scalar and $h$ is expected to arise at $O(\Pi^6)$, which gives an effect suppressed by $v^2/f^2$ compared to the $\Pi^4$ terms we considered here.)

We will study the potential in detail in the following section, but for now let us observe that generically $\lambda_{h\phi\phi}\sim\lambda$. Because experimentally $\lambda\approx0.13$ is small, we expect the new physics effect to be small as well. Consistently, we find that the corrections to the rates are entirely controlled by the modified $a,c$ couplings as soon as the NGBs are given a (universal) mass $m_\phi>100$ GeV. 

(It is in principle possible that the rate be enhanced compared to our estimates if the Higgs mixes with the singlet $\eta$, and the latter has a large anomalous coupling to photons~\cite{Montull:2012ik}. We will not study this possibility in the following.)

\begin{figure}
\begin{center}
\includegraphics[width=2.8in]{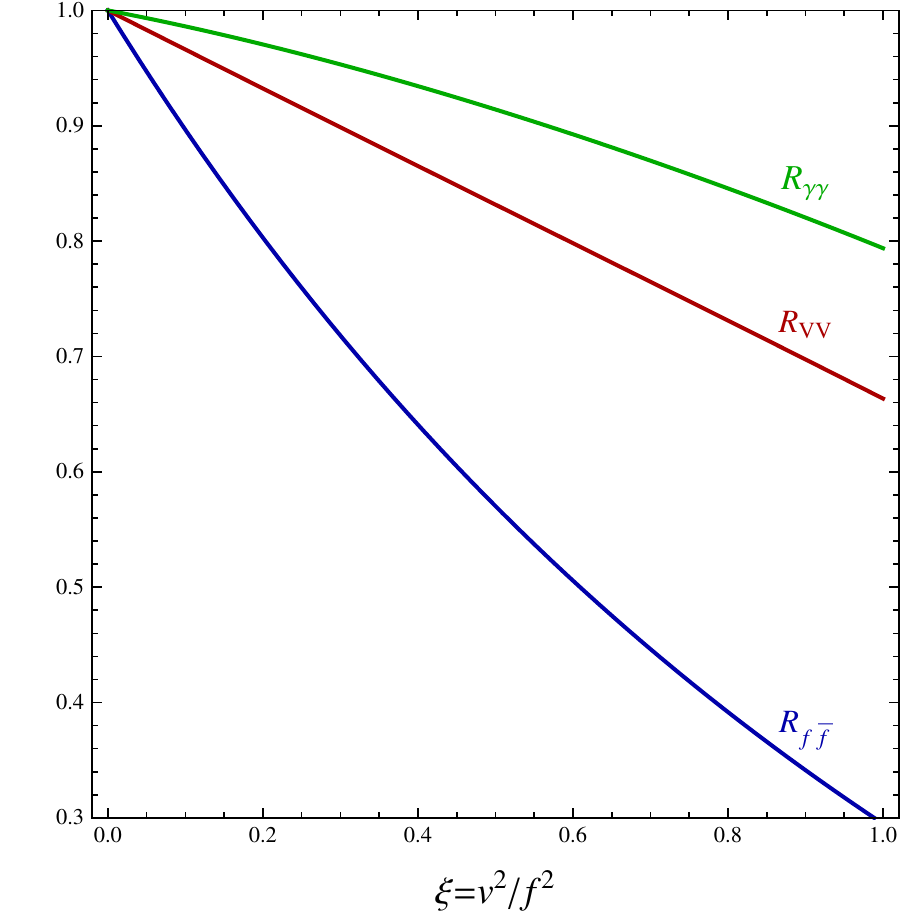}~~~~~~~~~
\caption{\small Signal strengths for the $\gamma\gamma,VV,f\overline{f}(gg)$ final states as a function of $\xi$. The model predicts a decay width for $h\to\gamma\gamma$ which is at most $10\%$ smaller than in the SM, a reduction always less then $20\%$ in $\Gamma(h\to VV)$, and finally a universal suppression of the couplings to fermions. In the figure we include the effect of the modified production rate (for the 7-8 TeV LHC) and total width, according to Eq.(\ref{R}). A fit to the current LHC data is presented in figure~\ref{FitXi} (see also figure~\ref{fitMCHM} right). 
\label{rates}}
\end{center}
\end{figure}

In conclusion, the couplings of the pseudo-NGB Higgs are the same of a fundamental Higgs boson up to a universal scaling given by $a,c$. The main effect of these modified couplings is a suppression of the Higgs production cross section times branching ratio compared to the SM:
\ba\label{R}
R_i&=&\frac{\sigma(pp\to h){\rm BR}(h\to i)}{\sigma_{\rm SM}(pp\to h){\rm BR}_{\rm SM}(h\to i)}\\\no
&\approx&\left(\frac{.88 c^2+.12 a^2}{.76 c^2+.24 a^2}\right)\times\frac{\Gamma(h\to i)}{\Gamma_{\rm SM}(h\to i)}.
\ea
In the second line we used the fact that the production at the LHC7-LHC8 and decay for $m_h\simeq126$ GeV are dominated respectively by gluon fusion ($\sim88\%$) and $h\to b\overline{b},\tau\overline{\tau},gg$ ($\sim76\%$), and that these rates scale as $c^2$ compared to the SM. While $\sigma(pp\to h)$ is always smaller than in the SM, the corresponding suppression of the Higgs width partially compensates this effect, so that $R_i\simeq\Gamma(h\to i)/\Gamma_{\rm SM}(h\to i)$ up to a $\lesssim10\%$ reduction. The numerical result for the ratios $R_{VV,\overline{f}f,\gamma\gamma}$ is shown in figure~\ref{rates}.

\begin{figure}
\begin{center}
\includegraphics[width=3.5in]{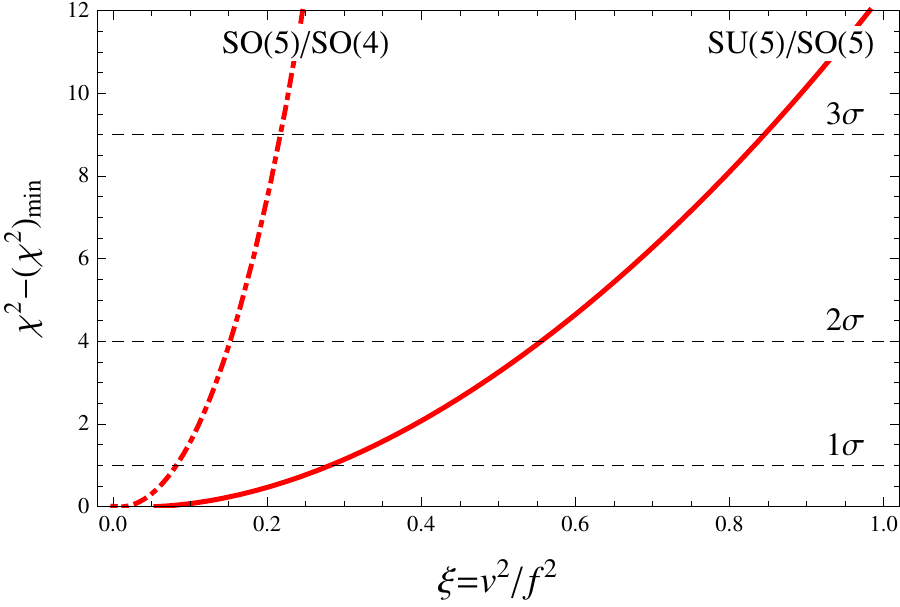}
\caption{\small Fit of the parameter $\xi\equiv v^2/f^2$ to the 2013 LHC Higgs data. See also figure~\ref{fitMCHM} (right).
\label{FitXi}}
\end{center}
\end{figure}

The Higgs boson couplings to the SM are essentially determined by a unique parameter $\xi$. We can therefore fit the 2013 LHC data with this single number. We again use the leading order approximation of $R_i$ employed in ref.~\cite{Giardino:2013bma}, and find excellent agreement with their full fit, consistently with their claim. The resulting $\chi^2$ is shown in figure~\ref{FitXi}. 

Remarkably, the $SU(5)/SO(5)$ model fits the data to better than $2\sigma$ as soon as $\xi\lesssim0.5$ ($f\gtrsim350$ GeV). We emphasize the present model does significantly better compared to the more optimistic line shown in~\cite{Giardino:2013bma} ($a=1, c=1-\xi$), because the $SU(5)/SO(5)$ values of $a,c$ are such to approach the ellipse in figure~\ref{fitMCHM} along the most favorable direction. For reference we also show the curve for the $SO(5)/SO(4)$ model of~\cite{Contino:2006qr} (red dot-dashed).

\section{The Scalar Potential}
\label{sec:Hpotential}

We now wish to discuss the various contributions to the Higgs potential $V_{\rm total}$. 

As argued in section~\ref{sec:picture}, the SM fermion contribution to $V_{\rm total}$ will have to involve the $Q$s, since there is no ``direct" coupling between the Higgs sector and $q,u$. We will first analyze the 1-loop contribution from the leading order Lagrangian~(\ref{EFT1}) (see section~\ref{sec:toploop}), and then in section~\ref{sec:quartic} consider the effects $\propto\widehat\epsilon_{f}^{~4}$ pointed out at the end of section~\ref{sec:picture}. Because the leading order contribution is small, the latter turn out to be in general {\emph{not}} negligible. We then discuss the gauge fields contribution in section~\ref{sec:gauge}, and a potential term from the constituents mass (see section~\ref{sec:mass}).

In section~\ref{sec:triplet} we will emphasize that new contributions to the mass of $\phi_+$ and the Higgs quartic are required to accommodate the EW data, and discuss their origin. 

The model will be shown to naturally explain the Higgs mass and the EW scale in section~\ref{EWSBnaturally}.

\subsection{SM fermions and gauge fields}

\subsubsection{A negative $\mu^2$ from top loops}
\label{sec:toploop}

Within the leading EFT in~(\ref{EFT1}) the dominant contribution to $V_{\rm total}$ is captured by the 1-loop Coleman-Weinberg potential. This can be immediately derived from the charge $2/3$ fermion mass matrix: 
\ba\label{M23}
M_{2/3}=\left( \begin{array}{cc}  m_QU&\lambda_u^\dagger \\
\lambda_q&0\end{array}\right).
\ea
On general grounds we expect the potential to be the trace of a function of the two invariants $\lambda_u^\dagger\lambda_u$ and $U\lambda_q^\dagger\lambda_qU^\dagger$. As anticipated, we thus see that the generation of a non-trivial potential requires both mixing parameters $\lambda_{q,u}$. In addition, we know that powers of $m_Q$ must be present, so the potential is actually finite at any order. The Coleman-Weinberg potential reads ($p_E$ is the Euclidean 4-momentum):
\ba\label{toppotential}
\delta V_{\rm IR}&=&-2N_c\int\frac{d^4p_E}{(2\pi)^4}{\rm tr}\left(p_E^2+M_{2/3}^\dagger M_{2/3}\right)\\\no
&=&-\frac{N_c}{16\pi^2}{\rm tr}\left[(M_{2/3}^\dagger M_{2/3})^2\log M_{2/3}^\dagger M_{2/3}\right]+{\rm const},
\ea
where in the last step we used the fact that the trace of $(M_{2/3}^\dagger M_{2/3})^2$ does not depend on the NGBs, which confirms the Coleman-Weinberg potential is actually finite. The subscript ``IR" is there to remind us that this contribution is controlled by scales $m_Q\ll m_\rho$.

Inspecting the structure of the spurion $\lambda_u$ and of $\Pi$, one finds that the fermion contribution does not generate any potential for $\phi_{+,0},\eta$ alone. This may be seen as a consequence of the quantum numbers of the NGBs, and specifically of the fact that there exists no non-derivative coupling involving only $\phi_{+,0},\eta$ at quadratic order in the SM fermions.

From the explicit expression of the determinant of $M_{2/3}$ given in Appendix~\ref{app:Mass} we see that $\delta V_{\rm IR}$ can be expressed as an expansion in powers of $(\epsilon_q\epsilon_u \sin)^2$. In practice the potential can be very well approximated by the leading $O(\sin^2)$ term. This is a consequence of the fact that $|\sin|<1$ and that in the limit in which one of the mixing angles is much smaller than the other, say $\epsilon_q\ll\epsilon_u$, the higher order terms in $\sin^2$ are suppressed by higher powers of the small parameter $\epsilon_q^2$. A numerical analysis of the exact expression~(\ref{toppotential}) confirms that retaining only the leading term $\propto \sin^2$ is a very accurate approximation (better than the percent level) throughout the parameter space ($0<\epsilon_{q,u}<1$). This allows us to drastically simplify our analysis by writing
\ba\label{topEFT}
\delta V_{\rm IR}\approx\alpha_{\rm IR} \sin^2,
\ea
with
\ba\label{alphat}
\alpha_{\rm IR}&=&-\frac{N_cy_t^2}{16\pi^2}f^2\frac{m_{Q_-}^2m_{Q_0}^2}{m_{Q_-}^2-m_{Q_0}^2}\log\frac{m_{Q_-}^2}{m_{Q_0}^2}<0.
\ea
Here $m_{Q_-}, m_{Q_0}$ are the physical masses of the exotic fermions $Q_{-,0}$ before EW symmetry breaking (see section~\ref{sec:collider} and Appendix~\ref{app:Mass}).

The above formula shows that the Higgs mass is naturally light provided at least one of the fermions is light. In the limit in which one of the masses $m_{\rm heavy}$ is much larger than the other the quadratic divergence is cutoff at a scale of order $m_{\rm light}^2\log m_{\rm heavy}^2/m_{\rm light}^2$ and the theory is still natural as long as $m_{\rm light}$ is not heavier than roughly 1 TeV (see figure~\ref{naturalness}, left). 

The structure of Eq.~(\ref{alphat}) is quite independent on the symmetry breaking pattern. A similar result follows for example from the $SO(5)/SO(4)$ version of the EFT~(\ref{EFT}), see~\cite{Matsedonskyi:2012ym}, or in any $SO(5)/SO(4)$ model in which a collective breaking mechanism is at work~\cite{Marzocca:2012zn}\cite{Pomarol:2012qf}. Our formula also agrees with~\cite{Berger:2012ec}, which considers a model based on $SU(3)/SU(2)$ in the decoupling limit $m_{\rm heavy}\to m_\rho$.

\subsubsection{A Higgs quartic from quark loops}
\label{sec:quartic}

The potential $\delta V_{\rm IR}$ has been calculated using the truncated Lagrangian~(\ref{EFT1}). In section~\ref{sec:picture} we have argued that this leading order Lagrangian neglects terms proportional to the couplings $\widehat\epsilon_{q,u}$, such as those in the second and third line in~(\ref{EFT2}). These were neglected in the previous sections because $\widehat\epsilon_{q,u}<\epsilon_{q,u}$, but cannot be neglected here!

The reason was already anticipated in section~\ref{sec:picture}: $\delta V_{\rm IR}$ is finite and dominated by physics at $\sim m_Q$, while the terms $\propto\widehat\epsilon_{q,u}$ in general result in UV divergent corrections to the Higgs potential, such that at the end of the day the $\widehat\epsilon_{q,u}/\epsilon_{q,u}$ suppression is compensated by an enhancement of a power of $m_\rho^2/m_Q^2$. For example, the second and third line in~(\ref{EFT2})~\footnote{Recall that the notation of Eq.~(\ref{EFT2}) is symbolic: the presence of derivative terms is implicitly understood.} give rise to a divergent potential scaling precisely as $\delta V_{\rm EFT}$, which in fact acts as a counterterm. The overall effect may be written as
\ba\label{UVpot}
\delta V_{{\rm UV},ff'}\sim\frac{N_c}{16\pi^2}m_\rho^2m_Q^2\left({\widehat\epsilon_f}^{~2}{\widehat\epsilon_{f'}}^{~2}+O({\widehat\epsilon_f}^{~6})\right).
\ea
We dubbed these corrections ``UV contributions", since they are saturated at scales of order $m_\rho$.

The UV contribution $\propto(\widehat\epsilon_q\widehat\epsilon_u)^2$ has the following spurion structure:  
\ba\label{A1violating}
\sim\frac{N_c}{16\pi^2}m_\rho^2m_Q^2{\rm tr}\left[\widehat\epsilon_u^\dagger\widehat\epsilon_u U \widehat\epsilon_q^\dagger\widehat\epsilon_qU^\dagger\right]\propto\sin^2+\,{\rm other~NGBs}.
\ea
In section~\ref{sec:picture} we saw this is expected to be parametrically comparable to the leading order~(\ref{topEFT}). Because it also has the very same functional dependence, it will be regarded as a perturbation of the ``IR potential".

Larger effects may arise at order $(\widehat\epsilon_u\widehat\epsilon_u)^2$ or $(\widehat\epsilon_q\widehat\epsilon_q)^2$. In the case $\lambda_q<\lambda_u$ the first type dominates while the latter can be neglected, and we find the following potential term:
\ba\label{A1invariant}
\sim\frac{N_c}{16\pi^2}m_\rho^2m_Q^2{\rm tr}\left[\widehat\epsilon_u^\dagger\widehat\epsilon_u U\Sigma_0\left(\widehat\epsilon_u^\dagger\widehat\epsilon_u U\Sigma_0\right)^*\right]\propto \sin^2+\,{\rm other~NGBs}+{\rm const},
\ea
This cannot be generated by the leading Lagrangian~(\ref{EFT1}), since $\Sigma_0$ does not appear there. Yet, this is the same function of $H$ as~(\ref{topEFT}), such that no qualitatively new effect is expected. Eq.~(\ref{A1invariant}) renormalizes the Higgs mass, and in a natural theory we would like this term be not too large.

We are therefore left with the regime $\lambda_q>\lambda_u$. In this case the main UV contribution to the potential is 
\ba\label{quartic}
\delta V_{{\rm UV},qq}&\supset&c_{\rm UV}\frac{N_c}{16\pi^2}m_\rho^2m_Q^2{\rm tr}\left[U\widehat\epsilon_q^\dagger\widehat\epsilon_q\Sigma_0\left(U\widehat\epsilon_q^\dagger\widehat\epsilon_q\Sigma_0\right)^*\right]\\\no
&\equiv&\gamma_q\left(1-\cos\right)^2+\,{\rm other~NGBs},
\ea
with $c_{\rm UV}$ of order unity. Eq.~(\ref{quartic}) gives a {\emph{quartic Higgs coupling}}. Its generalization including higher powers of $\widehat\epsilon_q^\dagger\widehat\epsilon_q$ contributes terms $O(1-\cos)^4$, which give $h^{4n}$ with $n>1$ but never a Higgs mass squared. We thus find that for $\lambda_q>\lambda_u$ the Higgs mass is dominantly generated by physics at scales $O(m_Q)$, while loops of $q$ saturated at $\sim m_\rho$ contribute a quartic. The two couplings are independent, and their relative magnitude is controlled by the otherwise undetermined ratio $\lambda_q/\lambda_u$. As $\lambda_q$ grows bigger than $m_Q$, $\gamma_q$ gets larger than $\alpha_{\rm IR}$, and the Higgs potential acquires a parametrically large quartic.~\footnote{It is worth stressing that a large $\gamma_q$ in natural models with $m_Q\gtrsim y_tf$ is achieved with $\epsilon_u\lesssim\epsilon_q$. This is important, because a very small $\epsilon_u/\epsilon_q$ would result in too large effects in $B_{d,s}$ and Kaon physics~\cite{KerenZur:2012fr}. In our framework the generic expectation from Eq.~(\ref{mf}) is in fact $\epsilon_u\gtrsim y_tf/m_Q$, where the lower bound is reached in the extreme limit $\lambda_q/\lambda_u\to\infty$.}

The presence of a Higgs quartic without mass term is simply the consequence of the collective breaking mechanism discussed in~\cite{ArkaniHamed:2002qy}. The coupling $\lambda_q^\dagger\lambda_q$ breaks $SU(5)$ down to its $SU(2)_L$ subgroup times an $SU(3)$ living in the lower-right corner of the generators. This latter acts non-linearly on the Higgs, such that no mass squared can be generated. Higher order terms in the potential are instead allowed because the shift symmetry can be compensated by the transformation of other NGBs, here $\phi_+$. At quadratic order in $\phi_+$ and quartic in the Higgs doublet, Eq.~(\ref{quartic}) reads
\ba\label{structure}
\delta V_{{\rm UV},qq}&\supset&\gamma_q~{\rm tr}\left[\left(2\frac{\phi_+}{f}+\frac{H\tilde H^\dagger}{f^2}+\dots\right)\left(2\frac{\phi_+}{f}+\frac{H\tilde H^\dagger}{f^2}+\dots\right)^\dagger\right]\\\no
&=&\frac{8\gamma_q}{f^2}~\frac{1}{2}{\rm tr}\,\phi_+^\dagger\phi_++\frac{\gamma_q}{f^4}(H^\dagger H)^2+\frac{2\gamma_q}{f^3}H^\dagger\phi_+\tilde H+{\rm hc}+\dots
\ea

The problem of Eq.~(\ref{quartic}) is that it violates custodial $SU(2)_R$ and $A_1$, and in particular it contains a trilinear $H^\dagger\phi_+\tilde H$ that generates a tadpole for $\phi_+$ and hence a vev for the EW triplet. This results in a tree-level correction to $T$ which would be unacceptably large in a model with large $\gamma_q$. We will discuss the impact of this effect, as well as a solution for this ``triplet problem", in section~\ref{sec:triplet}.

\subsubsection{One-loop potential from the gauge fields}
\label{sec:gauge}

The 1-loop potential from the gauge interactions may be parametrized by
\ba\label{gaugeEFT}
\delta V_{\rm gauge}&=&-c_{\rm g}\frac{m_G^2f^2}{16\pi^2}\left[g^2\sum_a{\rm tr}(T^a_LUT^a_LU^\dagger)+{g'}^2{\rm tr}(T^3_RUT^3_RU^\dagger)\right]\\\no
&=&-c_{\rm g}\frac{m_G^2f^2}{16\pi^2}\left[\frac{3g^2+g'^2}{2} (1+\cos)\right]+\,{\rm other~NGBs}\\\no
&=&-\beta_{\rm gauge}\cos+\,{\rm other~NGBs}+{\rm const},
\ea
where $m_G$ is the mass scale of the gauge boson partners (see section~\ref{sec:gauge1}) and the minus sign is conventional. Lattice simulations, as well as our experience with QCD, suggest that the gauge interactions tend to align the vacuum along the unbroken direction, so we expect $c_{\rm g}>0$ ($\beta_{\rm gauge}>0$). 

This potential also contributes to the masses of the other charged NGBs. Neglecting the hypercharge for simplicity, we find that $\phi_0,\phi_+$ pick up a common mass from the $SU(2)_L$ gauge loops: 
\ba
\delta m_{\phi_{+,0}}^2=c_g\frac{g^2}{4\pi^2}m_G^2.
\ea
This potential provides the only mass for $\phi_0$ so far, while a non-vanishing $m_\eta$ will have to wait until section~\ref{sec:mass}.

\subsection{A mass for the singlet}
\label{sec:mass}

Additional sources of $SU(5)$ breaking are required to give a mass to $\eta$, since none of the parameters considered so far break the shift symmetry acting on it. We therefore introduce the same spurion as in~\cite{Katz:2005au}, which may be interpreted as arising from a hypothetical mass term ${\cal M}\to g{\cal M}g^t$ for the UV constituents of the NGBs:
\ba\label{spurion}
\delta V_{\rm mass}=-m_\rho f^2{\rm tr}\left[{\cal M}\Sigma_0U^\dagger\right]+{\rm hc},
\ea
with
\ba\label{preonmass}
{\cal M}=\left( \begin{array}{ccc}  {\cal M}_- &~& \\
~&{\cal M}_+&\\\no
& &{\cal M}_0 \end{array}\right)\Sigma_0.
\ea
Here we allowed the most general mass term compatible with $SU(2)_L\times U(1)_Y$. We will show that the violation of the custodial symmetry obtained when ${\cal M}_+\neq {\cal M}_-$ is negligible in natural models (see section~\ref{sec:EWPT}). 

The above term respects the discrete symmetry $A_1$ if ${\cal M}$ is real, and gives 
\ba\label{betamass}
\delta V_{\rm mass}&=&-m_\rho f^2({\cal M}_++{\cal M}_-+2{\cal M}_0) \cos+\frac{3}{2}({\cal M}_++{\cal M}_-)+\,{\rm other~NGBs}\\\no
&=&-\beta_{\rm mass} \cos+\,{\rm other~NGBs}+{\rm const}.
\ea
$\beta_{\rm mass}$ provides the only mass term for $\eta$ in our model. ${\cal M}$ will also split the masses of the $Q$s at order ${\cal M}/m_\rho\ll1$.

\subsection{The Triplet problem and a new Higgs quartic}
\label{sec:triplet}

There are a number of neutral scalars in our model that could acquire a vev compatibly with QED: $h$, $\eta$, $\phi_0=\sigma^3 v_0/\sqrt{2}$ (with $v_0$ real), and finally $\phi_+=v_+\sigma_-$ (where $v_+$ may be complex). The most dangerous situation is when the electroweak triplets get a vev, which would generally result in a breaking of the custodial symmetry. The condition to preserve custodial $SU(2)_R$ is $v_+=-i\sqrt{2}v_0$, which requires a vev for the electric-neutral $CP$-odd component of $\phi_+$. 

However, $v_0$ and $v_+$ will not satisfy this simple relation because the potential receives important corrections from the custodial-violating coupling $\lambda_q$. Therefore, in general the potential $V_{\rm total}$ will contain a term $H^\dagger\phi_+\tilde H$ which leads to a tadpole for $\phi_+$, and eventually $v_+\neq0$, while the analogous operator involving $\phi_0$ is forbidden by the approximate CP invariance in Eq.~(\ref{A1}), and could only arise at a higher loop level. The vev $v_+$ will in turn contribute positively to the $Z^0$ mass at tree-level, and significantly affect the electroweak $T$ parameter:
\ba\label{T}
\alpha_{\rm em} \Delta T&=&-\frac{4v_+^2}{v^2}.
\ea 
The impact of this effect on the Little Higgs model of~\cite{ArkaniHamed:2002qy} was discussed in~\cite{Han:2003wu}\cite{Csaki:2003si}.

(The vev $v_+$ will also induce a mixing $h-\phi_+^{(0)}$. The mixing angle parametrically scales as $\sqrt{\alpha_{\rm em}T}$, and is always negligible in a realistic theory. For this reason we will focus on the $T$ parameter and ignore the effect of the mixing.)

The above trilinear coupling can only be generated by interactions that violate $A_1$. It is easy to see that both non-fermionic contributions~(\ref{gaugeEFT}) and~(\ref{spurion}) are $A_1$-invariant. The full potential $V_{\rm total}$ will be symmetric under $U\to U^\dagger$ only if combined with the spurionic transformation
\ba
\lambda_i^\dagger\lambda_i\to\Sigma_0(\lambda_i^\dagger\lambda_i)^*\Sigma_0.
\ea
Among the potential terms induced by the loops of the SM fermions, it turns out that only the operator~(\ref{A1invariant}) is invariant, because $\lambda_u^\dagger\lambda_u=\Sigma_0(\lambda_u^\dagger\lambda_u)^t\Sigma_0$ but the same is not true for $\lambda_q^\dagger\lambda_q$. One can explicitly verify that Eq.~(\ref{A1invariant}) does indeed contain only even powers of $\Pi$. Consistently with our symmetry argument we get
\ba\label{tri}
V_{\rm total}\supset\frac{2\gamma_q-\alpha}{f^3}H^\dagger\phi_+\tilde H+{\rm hc},
\ea
where $\alpha$ is understood to contain only the contribution of~(\ref{topEFT}) and~(\ref{A1violating}), whereas $\gamma_q$ is defined in~(\ref{quartic}) (see also Eq.~(\ref{structure})). The above trilinear leads to $v_+=-(2\gamma_q-\alpha) v^2/f^3m_{\phi_+}^2$, where the equality is correct up to $O(\xi)$.

To solve our problem we cannot decouple $\phi_+$ by taking $\gamma_q$ very large, as in so doing we also enhance $v_+$. We cannot increase $m_{\phi_+}$ by taking ${\cal M}$ large along the trajectory $2{\cal M}_0\approx-({\cal M}_++{\cal M}_-)$ (see Eq.~(\ref{betamass})), either. This typically requires some tuning and, even more importantly, it leads to a large vev for either $\phi_{+,0}$ or $\eta$, which then feeds back into the Higgs potential and leads to an unacceptably large Higgs mass.

The option  $\gamma_q\ll|\alpha|$ eliminates the quartic, but does not help. Indeed, in the absence of a quartic coupling one is forced to tune $\alpha_{\rm IR}$ and the other parameters of the potential, and this effectively requires a $\alpha_{\rm IR}$ parametrically larger than in a model with a Higgs quartic (this will be clearer from Eq.~(\ref{higgsmass1}) below, from which in the absence of the Higgs quartic one would find $\alpha/f^2\sim m_h^2/\xi$ and hence a too large $T$).

The bottom line is that one must increase the mass of the triplet without enhancing the Higgs mass nor the trilinear, and without giving up the quartic. Additional structure has to be invoked to accomplish this, but fortunately we have all the ingredients to understand how this can be done.

In~(\ref{structure}) we learnt an important lesson: {\emph{a large Higgs quartic is always accompanied by a large $m_{\phi_+}$}}. Finding a solution of the triplet problem is therefore equivalent to finding a new $A_1$-invariant Higgs quartic analogous to~(\ref{quartic}). This in general requires new physics below $m_\rho$ with couplings having transformation properties analogous to $\lambda_q^\dagger\lambda_q\propto\Delta_-^\dagger\Delta_-$. In particular, note that the following structure
\ba\label{structure'}
&&{\rm tr}\left[U\Delta_-^\dagger\Delta_-\Sigma_0\left(U\Delta_-^\dagger\Delta_-\Sigma_0\right)^*+U\Delta_+^\dagger\Delta_+\Sigma_0\left(U\Delta_+^\dagger\Delta_+\Sigma_0\right)^*\right],
\ea
with $\Delta_+=\left({\bf 0}_{2\times2}\quad{\bf 1}_{2\times2}\quad{\bf 0}_{2\times1}\right)$, is the $A_1$-invariant version of~(\ref{quartic}) and would do the job. In the following we will simply {\emph{assume}} the existence of the above $A_1$-symmetric spurion $\Delta_-^\dagger\Delta_-\otimes\Delta_-^\dagger\Delta_-+\Delta_+^\dagger\Delta_+\otimes\Delta_+^\dagger\Delta_+$. In Appendix~\ref{app:quartic} we will present an explicit model where it naturally appears. 

Armed with our new spurion one can construct the following $A_1$-symmetric potential
\ba\label{quartic'}
\delta V_{\rm UV}\supset\gamma_{\rm UV}(1-\cos)^2+\frac{8\gamma_{\rm UV}}{f^2}~\frac{1}{2}{\rm tr}\,\phi_+^\dagger\phi_++\cdots.
\ea
The trilinear is not generated, and the vev of the triplet will be under control for sufficiently large $\gamma_{\rm UV}$ (see section~\ref{sec:fit}).

We conclude by observing that, similarly to what has been done above, the authors of~\cite{Agashe:2004rs} added a spurion transforming under a $4\otimes4\otimes4\otimes4$ of $SO(4)\sim SU(2)_L\times SU(2)_R$ to their minimal $SO(5)/SO(4)$ model in order to get an independent quartic. In the present framework the spurions necessary to generate a Higgs quartic transform analogously.

\subsection{EW Symmetry Breaking, Naturally}
\label{EWSBnaturally}

Summing up all the terms found in the previous subsections, the Higgs potential may be written as follows
\ba\label{totalpot}
V_{\rm total}&=&\alpha \sin^2-\beta \cos+\,\gamma(1-\cos)^2+\,{\rm other~NGBs}.
\ea
The scalars $\eta,\phi_0$ do not get a vev and so have no role in EW symmetry breaking, whereas $\phi_+$ brings $O(v_+^2/v^2)$ corrections that are negligible in realistic models, see Eq.~(\ref{T}). 

In~(\ref{totalpot}), $\alpha$ receives contributions from Eq.~(\ref{topEFT}) and from the UV-sensitive terms in~(\ref{A1violating}) and~(\ref{A1invariant}), whereas Eqs.~(\ref{gaugeEFT})(\ref{spurion}) enter $\beta$, and finally $\gamma=\gamma_q+\gamma_{\rm UV}$ is the sum of the UV-sensitive effects from loops of the SM quark doublet~(\ref{quartic}) and the $A_1$-symmetric spurion introduced above~(\ref{quartic'}) and discussed in detail in Appendix~\ref{app:quartic}. Physics below the cutoff unambiguously gives a negative $\alpha$ (as well as a positive contribution to $\beta$). The remaining effects depend on the sign of unknown coefficients.

 Expanding in powers of the Higgs boson $h$, we find that $\mu^2=(2\alpha+\beta)/f^2$ plays the role of a Higgs bare mass, and $\lambda=(-8\alpha-\beta+6\gamma)/6f^4$ of the Higgs quartic coupling. Crucially, these are controlled by distinct parameters.

A straightforward analysis of the potential reveals that for $\gamma>\alpha$ and $2\alpha-4\gamma\leq\beta\leq-2\alpha$ the absolute minimum of $V_{\rm total}$ is at $\cos=(\gamma+\beta/2)/(\gamma-\alpha)$, that is
\ba\label{xi}
\xi=\frac{-2\alpha-\beta}{\gamma-\alpha}.
\ea
In order to have a fully realistic model with $\xi<1$ we just need $\alpha+\beta+\gamma>0$. This latter condition plus those above~(\ref{xi}) are equivalent to 
\ba\label{conditions}
\left\{ \begin{array}{c}  \beta+2\alpha\leq0 \\
\alpha+\beta+\gamma>0. \end{array}\right.
\ea
These relations are generic, and $\xi<1$ is natural. For example, if $\beta$ can be neglected, they simply read $\alpha\leq0,\gamma>|\alpha|$. In this case, $\xi$ ranges between $0.3-0.6$ for $\gamma/|\alpha|\in[2,6]$.

Let us now see if the model can also explain the value of the physical Higgs boson mass without fine-tuning. From $V_{\rm total}$ we find 
\ba\label{higgsmass1}
m_h^2&=&2(\gamma-\alpha)\frac{\sin^2}{f^2}\\\no
&=&2\frac{|2\alpha+\beta|}{f^2}\left(1-\frac{\xi}{4}\right).
\ea
where in the second step we used~(\ref{xi}). It turns out that $m_h^2$ depends only weakly on $\gamma$ and satisfies $0\leq m_h^2\leq-2\mu^2$, where the upper bound is approached in the limit of large $\gamma$. (For ease of clarity, we mention that the regime $|\gamma|\gg|\alpha|$ does not correspond to the large quartic limit $\lambda\gg1$, which is instead defined by $|\gamma|\gg f^4$.)

For completeness let us here summarize the masses of the other NGBs:
\ba\label{NGBmass}
m_{\phi_{0}}^2&=&c_g\frac{g^2}{4\pi^2}m_G^2+2m_\rho({\cal M}_++{\cal M}_-)\\\no
m_{\phi_{+}}^2&=&c_g\frac{g^2}{4\pi^2}m_G^2+2m_\rho({\cal M}_++{\cal M}_-)+\frac{8\gamma}{f^2}\\\no
m^2_\eta&=&m_\rho\frac{2{\cal M}_++2{\cal M}_-+16{\cal M}_0}{5},
\ea
with $\gamma=\gamma_q+\gamma_{\rm UV}$. We see that $m_{h,+,0,\eta}^2$ are in practice independent parameters. For $|{\cal M}_{+,-,0}|m_\rho\ll c_{\rm g}g^2 m_\rho^2/4\pi^2$ and $\gamma>0$ we expect $m_\eta^2\ll m_{\phi_0}^2<m_{\phi_+}^2$, with the mass of $\phi_0$ in~(\ref{NGBmass}) comparable to $m_h$ in any natural theory.

The Higgs mass depends on the incalculable parameters $\alpha,\beta$. Now, we have seen that there are two, potentially comparable effects contributing to $\alpha$. First, there is a calculable effect we called $\alpha_{\rm IR}$, and second, an unknown effect from UV physics, see Eq.~(\ref{A1violating}). To determine whether or not $m_h=126$ GeV is a realistic prediction of the model we thus proceed as follows. We allow a partial accidental cancellation of order $\sim1/\Delta$ between the two terms entering $\alpha$, as well as these and $\beta$, and define
\ba\label{higgsmass}
m_h^2&\equiv&\frac{4\left|\alpha_{\rm IR}\right|}{f^2}\times\frac{1}{\Delta},
\ea
with $\alpha_{\rm IR}$ defined in~(\ref{alphat}). Here $\Delta$ should give a measure of the fine-tuning necessary to get the right Higgs mass. But according to the claims below Eq.~(\ref{conditions}), this in fact represents the {\emph{total}} tuning as defined in~(\ref{deftuning}).

\begin{figure}
\begin{center}
\includegraphics[width=2.7in]{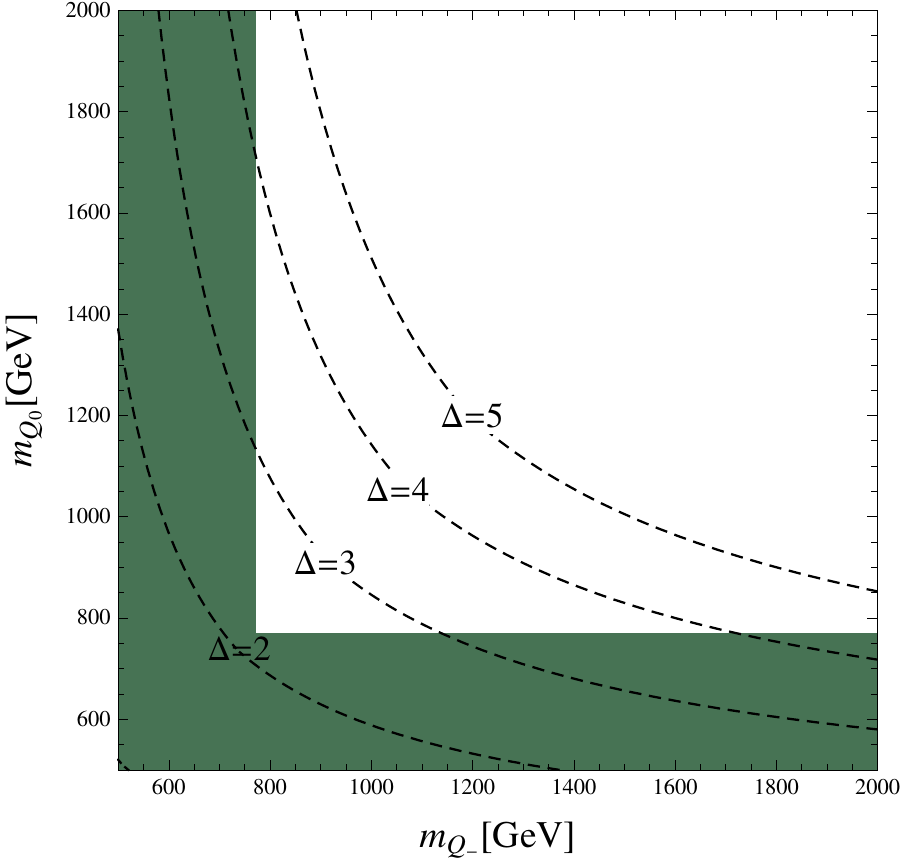}~~~~~~~~~~~~\includegraphics[width=2.7in]{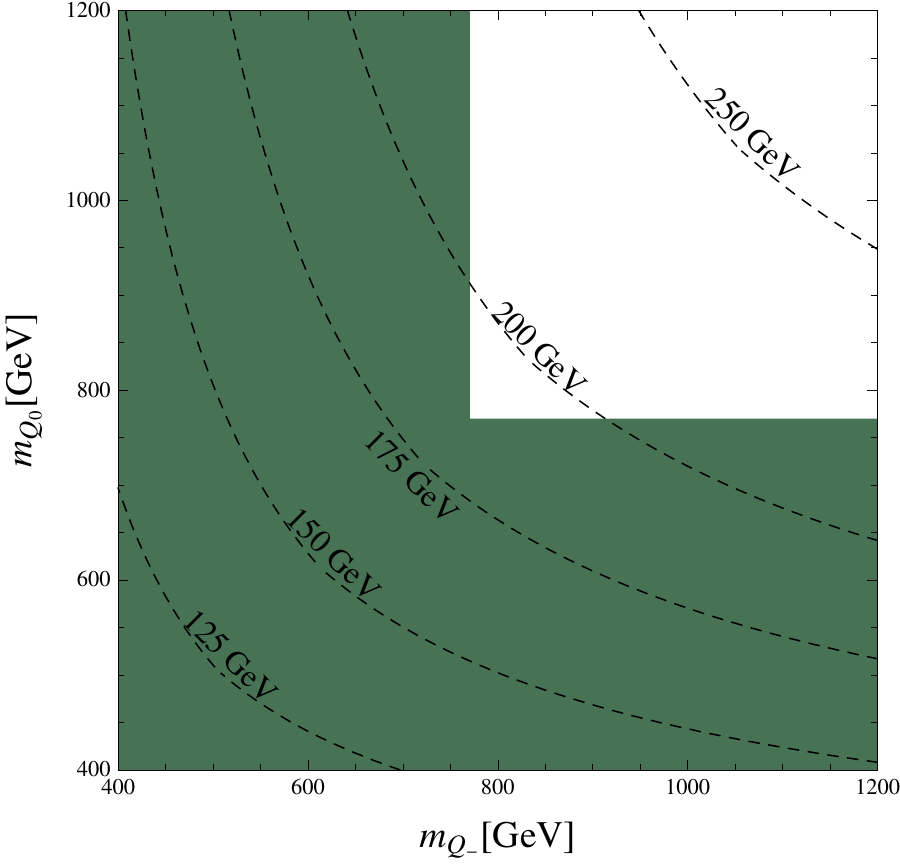}
\caption{\small Left: Contours of the fine-tuning parameter $\Delta=|4\alpha_{\rm IR}|/(126$ GeV$)^2f^2=2,3,4,5$ as a function of the masses of the exotic fermions $Q_{-,0}$, see~(\ref{alphat}). Right:  Contours of constant $|4\alpha_{\rm IR}/f^2|^{1/2}$. The green area below $m_{Q_-,Q_0}=770$ GeV is currently probed by direct searches at the LHC (see section~\ref{sec:collider}).
\label{naturalness}}
\end{center}
\end{figure}

Contours of constant $\Delta$ are plotted in figure~\ref{naturalness} (left) for $m_h=126$ GeV and $y_t$ renormalized at the scale max$(m_{Q_-},m_{Q_0})$. The expected Higgs boson mass for $\Delta=1$ is plotted in the right figure. The green area defined by $m_{Q_-,Q_0}\lesssim770$ GeV represents the region of parameter space currently probed by direct collider searches on the top partners, although we will argue in section~\ref{sec:collider} that the current constraints do not always apply to the present model. We will think of the filled region in figure~\ref{naturalness} as ``excluded", but keep in mind this conclusion is conservative.

Taking the bound at face value we see that the Higgs boson mass is naturally found at about $m_h\gtrsim200$ GeV. Given the discussion in section~\ref{sec:fermion} we expect this to be a model-independent implication of NGB Higgs models. In order to accommodate the observed value a cancellation of the order of one part in $2-3$ is sufficient (i.e. $\Delta\gtrsim2$).


We conclude pseudo-NGB Higgs models based on $SU(5)/SO(5)$ offer enough freedom to parametrically account for both the smallness of the weak scale {\emph{and}} $m_h=126$ GeV with virtually no tuning.

\section{Collider Signatures}
\label{sec:collider}

Before EW symmetry breaking $P_LQ_-$ mixes with $q$ and $P_RQ_0$ with $u$. As a result the masses of these fields are renormalized, while the mass of $Q_+$ stays $m_Q$ (modulo possible model-dependent corrections). The mass matrix can be easily diagonalized by the following redefinitions:
\ba
\left( \begin{array}{c}  
P_LQ_-\\
q
\end{array}\right)\to
\left( \begin{array}{cc}  
\sqrt{1-\epsilon_q^2}&~~-\epsilon_q\\
\epsilon_q&~~\sqrt{1-\epsilon_q^2}
\end{array}\right)
\left( \begin{array}{c}  
P_LQ_-\\
q
\end{array}\right),
\ea
where $\epsilon_{q,u}$ are defined in Eq.~(\ref{epsilonq}), and similarly
\ba
\left( \begin{array}{c}  
P_RQ_0\\
u
\end{array}\right)\to
\left( \begin{array}{cc}  
\sqrt{1-\epsilon_u^2}&~~-\epsilon_u\\
\epsilon_u&~~\sqrt{1-\epsilon_u^2}
\end{array}\right)
\left( \begin{array}{c}  
P_RQ_0\\
u
\end{array}\right).
\ea

Replacing the rescaled fields back in the Lagrangian~(\ref{EFT1}) we get
\ba\label{EFTQ}
{\cal L}_{\rm EFT}&=&{\cal L}_{\rm kin}+m_{Q_-}\overline{Q_-}Q_-+m_{Q_+}\overline{Q_+}Q_++m_{Q_0}\overline{Q_0}Q_0\\\no
&+&\frac{m_Q}{f}\epsilon_q\epsilon^\dagger_u\overline{q}\tilde Hu+{\rm hc}\\\no
&+&\frac{m_Q}{f}\epsilon_q\overline{q}\left[-i\left(\phi_0+\frac{\eta}{\sqrt{10}}\right)Q_--\phi_+^\dagger Q_++\tilde HQ_0+{\rm hc}\right]\\\no
&+&\frac{m_Q}{f}\left[-\overline{Q_-}\tilde H-\overline{Q_+}H+4i\overline{Q_0}\frac{\eta}{\sqrt{10}}+{\rm hc}\right]\epsilon^\dagger_uu\\\no
&+&\dots,
\ea 
where the dots stand for terms $O(\epsilon_{q,u}^2)$, those not involving the SM fields, as well as terms with more powers of the NGBs, and 
\ba
m_{Q_-}&=&\sqrt{\lambda_q\lambda_q^\dagger+m_Q^2}~~~~~~~~~~~~~m_{Q_+}=m_Q~~~~~~~~~~~~m_{Q_0}=\sqrt{\lambda_u\lambda_u^\dagger+m_Q^2}.
\ea
The masses of the various components of the $Q$s will get split after electroweak symmetry breaking (see Appendix~\ref{app:Mass}).

\subsection{The lightest Exotic Fermion}

We saw that the lightest exotic fermion is typically $Q_+$, transforming as a $(3,2)_{7/6}$ of the SM gauge group $SU(3)_c\times SU(2)_L\times U(1)_Y$.  This splits into a charge $5/3$ plus a charge $2/3$ after the Higgs gets a vacuum expectation value (vev). 

The decays of the $Q_+$ can be seen from~(\ref{EFTQ}). Importantly, the branching ratios into $W_L^\pm,Z_L^0$ involve right handed quarks and are proportional to $\epsilon_u^2$, while those into the exotic NGBs are $\propto\epsilon_q^2$, so these modes are controlled by independent parameters.

\begin{figure}
\begin{center}
\includegraphics[width=3.15in]{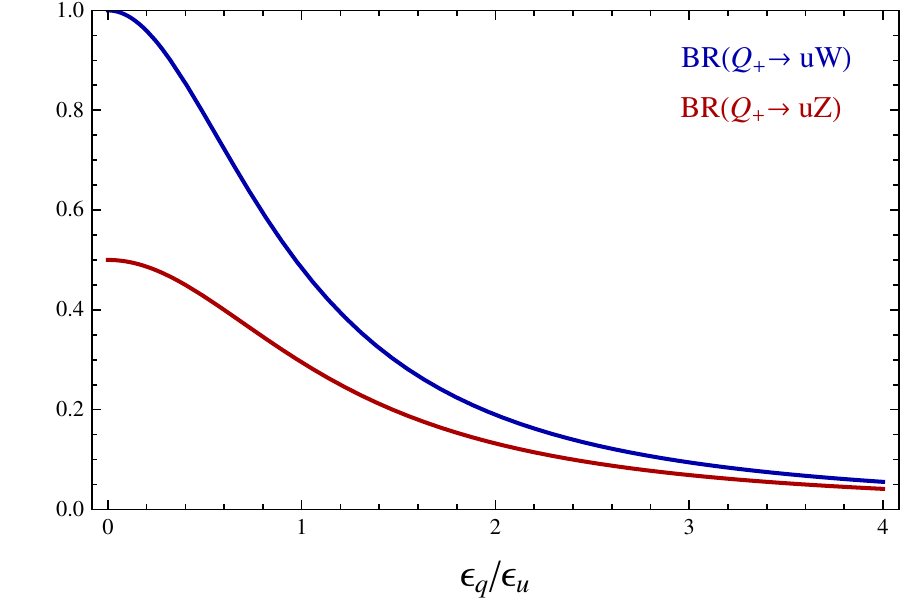}
\caption{\small Branching ratios of the two components of $Q_+$ into right-handed up-type quarks and longitudinal vector bosons as a function of the ratio $\epsilon_q/\epsilon_u$. In the figure we took $m_{Q_+}=800$ GeV and $m_{\phi_+}=500$ GeV.
\label{BR}}
\end{center}
\end{figure}

To see the impact of this feature, in figure~\ref{BR} we plot the branching ratios of $Q_+^{(5/3,2/3)}$ into $W_L^\pm,Z_L^0$ as a function of the ratio $\epsilon_q/\epsilon_u$ for the case $m_{\phi_+}=500$ GeV and $m_{Q_+}=800$ GeV. We see that the BRs into longitudinal gauge bosons quickly become subdominant when $\epsilon_q\gtrsim \epsilon_u$. Similarly, single production is suppressed. Therefore, the charge-$5/3$ component will be pair produced and then decay $\sim66\%$ of the times into $\phi_+^{(++)}d_{Li}$, and the remaining into $\phi_+^{(+)}u_{Li}$, whereas $Q_+^{(2/3)}$ will produce $\phi_+^{(+)},\phi_+^{(0)}$ plus left handed quarks.

We thus learn that for $\lambda_q>\lambda_u$ the phenomenology of the lightest top partner depends crucially on the decay modes of $\phi_+$, and may potentially depart from that of the minimal $SO(5)/SO(4)$ scenario~\cite{Contino:2008hi}\cite{Mrazek:2009yu} (effectively recovered in the limit $\epsilon_q/\epsilon_u\ll1$ of figure~\ref{BR}).

An important role in the determination of the appropriate search strategy is played by the doubly charged component $\phi^{(++)}_+$. In models where $A_1$ is broken, the scalar mainly decays into $W^\pm W^\pm$ via the potential operator $H^\dagger\phi_+\tilde H$. In this case the signature of the $Q_+^{(5/3)}$ would typically be $W^+W^+b$ (assuming the decays into the heavy generations dominate), that is the same as if the $\phi_+$ was not there. Using $\sim20$/fb of data from the 8 TeV LHC, CMS found a bound of $m_{Q_+}\gtrsim770$ GeV for a $100\%$ branching ratio into $W^+t$~\cite{CMS}. One can consider alternative scenarios where $A_1$ is a good symmetry of the scalar potential. In this scenario the width of $\phi_+$ is controlled by the Yukawa operator in the third line of~(\ref{EFTbelowQ}), namely the flavor-diagonal operator $y_i\overline{q_i}\phi_+^\dagger H u_i/f$. This latter leads to $\phi^{(++)}_+\to W^+t_R\overline{b_L}$, which may be seen as occurring via an off-shell $Q_+$. The resulting process is now $Q_+^{(5/3)}\to W^+tb\overline{b}$, that is the same as above plus some additional b-jets.

The signature of the exotic state $Q_+$s would qualitatively change in non-minimal scenarios where the decays of $\phi_+$ into light jets happen to dominate, or when new couplings for $Q_+$ are turned on (as in the model of Appendix~\ref{app:quartic}). 
\subsection{Cascade decays}

The decays of the heavier $Q_-,Q_0$ may occur at zeroth order in the mixing parameters $\epsilon_{q,u}$ via the cascade decays:
\ba
Q_0\to H^\dagger Q_+~~~~~~~~~~~~~~~~~Q_-\to \phi_+^\dagger Q_+.
\ea
These modes generically dominate over the ones into SM fermions as soon as they are kinematically accessible.

With $m_{Q_0}>m_Q+m_W$ we have BR$(Q_0\to W^-Q_+^{(5/3)})\sim50\%$, which typically results in a final state with hard leptons (including $\ell^\pm\ell^\pm$ pairs), jets, and missing energy. We may thus conservatively require $m_{Q_0}\gtrsim770$ GeV as above. In more elaborate models with non-standard $Q_+$ decays, one expects a much lower bound.

The doublet $Q_-$ has the same quantum numbers as the SM quark doublet, and contains a pair of electric charges $-1/3,2/3$. The cascade decays $Q_-\to\phi^\dagger_+Q_+$ will include the doubly charged scalar, so a conservative bound of $m_{Q_-}\gtrsim770$ GeV may be imposed here as well (again with the caveat discussed at the end of the previous subsection). A generic possibility is that $Q_-\to\phi^\dagger_+Q_0$ is kinematically closed, since $\phi_+$ is the heaviest NGB. In this regime we find that, analogously to the lightest exotic, the $Q_-$ will mainly decay into left-handed SM fermions and NGBs when $\lambda_q>\lambda_u$. Again, single production of $Q_-$, being controlled by $\epsilon_u$, is suppressed in this regime, such that the dominant production mechanism will be pair production. Importantly, a crucial difference from the $Q_+$ case is that $Q_-$ decays will involve the scalars $\phi_0,\eta$. These are expected to have masses comparable to (or lower than) the Higgs boson, and decay into light jets via the operators in the fourth line of~(\ref{EFTbelowQ}). The $Q_-^{(-1/3)}$ will thus dominantly result in $t\overline{c}b$ or $b\overline{b}b$ final states (again assuming that decays into heavier states are favored), whereas $Q_-^{(2/3)}\to tb\overline{b}, cb\overline{b}$. A relevant constraint here follows from a lepton plus $\geq6j$ search ($2$ or more jets required to be $b$s) performed by the ATLAS collaboration with $\sim14/$fb of data at 8 TeV~\cite{ATLAS}.

\section{Electroweak Precision Tests}
\label{sec:EWPT}

We now discuss the various contributions to the EW $S$ and $T$ parameters and the process $Z^0\to b\overline{b}$, and finally present a fit in section~\ref{sec:fit}.

\subsection{The $S$-Parameter}
\label{sec:EWfit}

The $S$ parameter arises from:
\ba
{\cal L}_{\rm EFT}&\supset&-c_S gg'\frac{f^2}{m_\rho^2}W^a_{\mu\nu}B_{\mu\nu}~{\rm tr}\left(T_L^aUT_R^3U^\dagger\right)\\\no
&=&-c_S gg'\frac{v^2}{m_\rho^2}W^3_{\mu\nu}B_{\mu\nu}+\dots
\ea
This operator is generated at tree-level with an order one coefficient $c_S$ by physics at the scale $m_\rho$. The $S$ parameter is defined as
\ba\label{S}
\Delta S_{\rm UV}&=&c_S~16\pi\frac{v^2}{m_\rho^2}.
\ea
In the following we will use $c_S=1/2$, which is what one approximately finds in calculable 5D models, or by simply integrating out the first two vector resonances taken conservatively to be degenerate (see for instance~\cite{Contino:2006nn}). 

The light fields contribute at loop-level, and give just a small correction which we will neglect.

\subsection{The $T$-Parameter}

$T$ receives non-negligible contributions from both physics above the cutoff and the EFT. We start with the effect of the modified Higgs coupling to $W^\pm$~\cite{Barbieri:2007bh}, which leads to:
\ba
\Delta T_h&=&-\frac{3}{8\pi\cos^2\theta_w}(1-a^2)\log\left(\frac{m_\rho}{m_h}\right)=-\frac{3}{32\pi c_w^2}\xi\log\left(\frac{m_\rho}{m_h}\right)
\ea
with $a=\sqrt{1-\xi/4}\leq1$. 

The vev of the triplet gives a negative correction as well. In the scenario we discussed in section~\ref{sec:triplet}, and for $|\gamma_q|\lesssim|\alpha|<|\gamma_{\rm UV}|$, we find $v_+/v\approx\alpha \sqrt{\xi}/f^2m_{\phi_+}^2\approx \sqrt{\xi}m_h^2/4m_{\phi_+}^2$, where in the last step we made use of~(\ref{higgsmass1}). Eq.~(\ref{T}) finally reads
\ba\label{Tphi2}
\alpha_{\rm em} \Delta T_\phi=-c_\phi~\frac{\xi}{4}\left(\frac{m_h}{m_{\phi_+}}\right)^4
\ea
for some $c_\phi=O(1)$. At loop-level the exotic NGBs contribute negligibly.

The custodial symmetry is also broken by $\lambda_q$. One-loop diagrams of the $Q$s and the SM fermions must contain at least 4 powers of this coupling to give a non-vanishing $\alpha_{\rm em}\Delta T_{\rm Q}$. The analogous contribution involving fields of masses $m_\rho$ is always suppressed by powers of $m_Q^2/m_\rho^2$ and can be neglected. 

We will see in section~\ref{sec:fit} that a realistic model needs a positive $\Delta T_Q$. A dedicated analysis of a similar scenario revealed that the requirement $\Delta T_Q>0$ selects the region of parameter space $\lambda_q>\lambda_u$~\cite{Anastasiou:2009rv}\cite{Gillioz:2008hs}. In the extreme limit $\lambda_q\gg\lambda_u$ the calculation simplifies and, in a leading log approximation, we find
\ba\label{positiveT}
\Delta T_Q\sim\frac{N_c}{16\pi s_w^2}\frac{m_t^2}{m_W^2}\epsilon^2_q~\xi~\log\frac{m_Q^2}{m_t^2}.
\ea
When $\epsilon_u\lesssim\epsilon_q$ additional negative corrections exist that partially compensate~(\ref{positiveT}). Thus, Eq.~(\ref{positiveT}) may be viewed as an upper bound on how positive $T$ can get. In section~\ref{sec:fit} we will take $\Delta T_Q\sim0.1$, which is a perfectly realistic assumption.

Other sources of custodial symmetry breaking from loops involving the hypercharge and the heavy fields scale as $\alpha_{\rm em}\Delta T_{g}\sim\xi g'^2/16\pi^2$, and are thus subleading.

Finally, there might be additional corrections to $T$ if the strong dynamics violates $SU(2)_R$, for example if ${\cal M}_+\neq {\cal M}_-$ in~(\ref{preonmass}). In that case we can write a term~\footnote{${\cal M}$ is a triplet under the spurionic $SU(2)_R$, so we need at least two insertions to correct the mass term of the vector bosons. This term is analogous to a bulk mass for the dual 5D gauge symmetry, as the one discussed in~\cite{Agashe:2003zs}.} 
\ba
{\cal L}_T=c_T\frac{f^2}{m_\rho^2}~{\rm tr}\left[{\cal M}(D_\mu\Sigma)^*{\cal M}(D_\mu\Sigma)^*\right]+{\rm hc}
\ea
where $c_T=O(1)$ by NDA. We get
\ba\label{cT}
\alpha_{\rm em}\Delta T_{\rm UV}&=&\frac{c_T}{2}~\xi\left(\frac{{\cal M}_+-{\cal M}_-}{m_\rho}\right)^2\left[1+O(c_T{\cal M}^2/m_\rho^2)\right].
\ea
Loops of the $Q$s with insertions of ${\cal M}$ will contribute, as well, but these corrections are down by a factor $O((m_Q/f)^2N_c/16\pi^2)$ because of the partial composite nature of the fermions, and can safely be neglected. To estimate~(\ref{cT}), we assume that ${\cal M}_{+,-,0}$ have approximately the same magnitude. We further use the fact that ${\cal M}$ should give a small correction to the Higgs mass, that is $m_h^2\gtrsim2|{\cal M}_{+,-,0}|m_\rho$. This gives $\alpha_{\rm em}\Delta T_{\rm UV}\lesssim c_T\xi m^4_{h}/8m_\rho^4$, which is always negligible.

\subsection{$Z^0\to b\overline{b}$}

In our effective field theory~(\ref{EFT}) there is no mass for the bottom quark. The most straightforward way to remedy this is adding a second multiplet of $Q^{(d)}$s with $U(1)_X$ charges $-1/3$, in exact analogy with what happens in the up sector. The component with EW charge $2_{1/6}$ will couple to $q$ with a coupling $\lambda_q^{(d)}$, and the singlet to $d$ with coupling $\lambda_d$. Together they will determine the mass of the $b$ quark as in~(\ref{mf}), $m_b\propto\lambda_q^{(d)}\lambda_d^\dagger$.

Once $Q^{(d)}$ is included we find a tree-level correction to the $b_L$ coupling to $Z^0$~\footnote{Technically, the correction in~(\ref{db}) is induced because $\lambda_q^{(d)}$ violates the parity symmetry $P_{LR}\subset O(4)$ enforcing $\delta g_{b_L}=0$ at leading order in the mixing and at zero momentum~\cite{Agashe:2006at} (see also~\cite{Mrazek:2011iu}).}
\ba\label{db}
\delta g_{b_L}=\frac{1-\langle \cos\rangle}{2}|\epsilon_q^{(d)}|^2=\frac{\xi}{4}|\epsilon_q^{(d)}|^2,
\ea
where we defined  
\ba
\frac{g}{\cos\theta_w} Z_\mu~\overline{b_L}\gamma^\mu b_L \left(-\frac{1}{2}+\frac{1}{3}\sin^2\theta_w+\delta g_{b_L}\right).
\ea
Because $\lambda_q^{(d)}$ must be small in order to explain the small mass of the bottom quark, the experimental bound on $\delta g_{b_L}$ can easily be satisfied.

Corrections to $\delta g_b$ proportional to the larger coupling $\lambda_q$ arise at one-loop and require at least four insertions of $\lambda_{q,u}$~\cite{Agashe:2006at}. These can be parametrized below $m_Q$ by dimension-6 operators of the form $\overline{q}\gamma^\mu qH^\dagger D_\mu H$:
\ba\label{dbb}
\delta g_{b_L}&\sim&\frac{(m_Q/f)^4}{16\pi^2}\epsilon_q^2\epsilon_{q,u}^2\frac{v^2}{m_Q^2}\sim \frac{y_t^2}{16\pi^2}\xi~{\rm max}\left(1,\frac{\epsilon_q^2}{\epsilon_{u}^2}\right).
\ea
As long as $m_Q/m_\rho<1$, we anticipate only small corrections from physics above $m_\rho$. 

In the regime $\epsilon_q>\epsilon_u$ the positive $\Delta T_Q$ from $Q$-loops (see Eq.~(\ref{positiveT})) is typically correlated with a positive contribution to~(\ref{dbb}). This was pointed out in~\cite{Anastasiou:2009rv}\cite{Gillioz:2008hs} for an analogous model based on the coset $SO(5)/SO(4)$. The main qualitative difference between that model and the present framework is the effect of the exotic NGBs. However, it turns out that the 1-loop contribution of $\phi_{+,0},\eta$ to $\delta g_b$ is suppressed when $\epsilon_u<\epsilon_q$, so the above statement applies essentially unaffected to our model. This correlation, plus the experimental constraint on $\delta g_b$, result in an upper bound on $\Delta T_Q$. That bound was somewhat problematic in~\cite{Anastasiou:2009rv}\cite{Gillioz:2008hs} because in that model a good EW fit required a rather large $\Delta T_Q\gtrsim+0.3$. On the other hand, we will see in section~\ref{sec:fit} that in the present framework a smaller $\Delta T_Q$ suffices because of the smaller $\Delta T_h$ characterizing $SU(5)/SO(5)$. In this case the bound on $\delta g_b$ poses no serious constraint.

Momentum suppressed corrections to $Z^0\to b\overline{b}$ can be described by higher dimensional operators such as $D_\nu F_{\mu\nu}\overline{Q}\gamma^\mu Q$ generated at $\sim m_\rho$ (see for example~\cite{Panico:2011pw}). Estimating the coefficient according to NDA, accounting for the mixing with the SM quark doublet, and finally renormalizing the effect down to the $Z^0$ pole, these approximately give
\ba
\delta g_{b_L}&\sim&\frac{m_Q}{m_\rho}\left(\frac{m_Z}{m_\rho}\right)^2\epsilon_q^2.
\ea
This is expected to be well within experimental bounds.

\subsection{Fit of the Oblique Parameters}
\label{sec:fit}

\begin{figure}
\begin{center}
\includegraphics[width=2.7in]{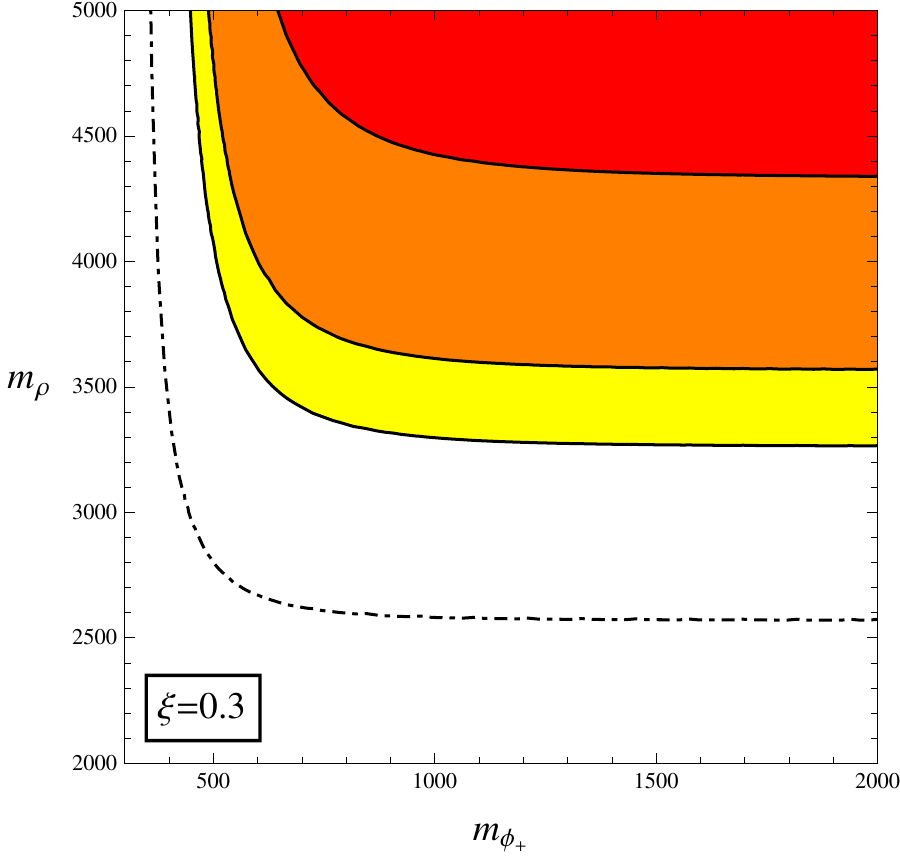}~~~~~~~~\includegraphics[width=2.7in]{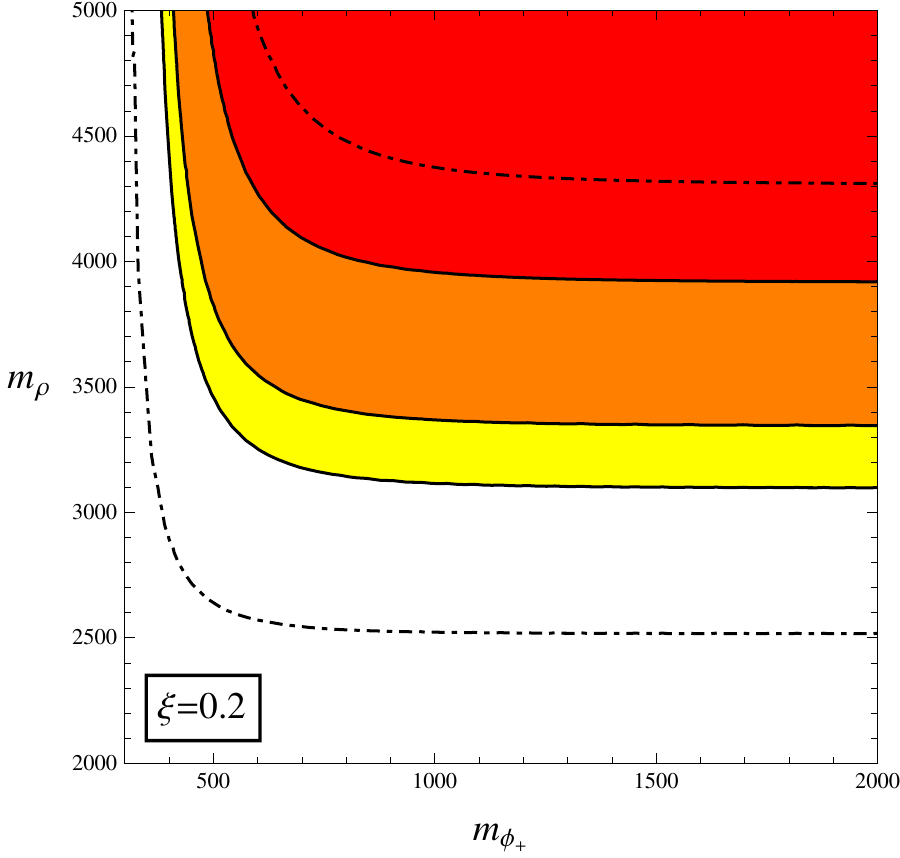}
\caption{\small $68,95,99\%$ CL fit of the parameters $S,T$ ($U=0$) as a function of the Higgs compositeness scale $m_\rho$ and $m_{\phi_+}$ for $\xi=0.2$ (left) and $\xi=0.3$ (right). In each plot we took $c_S=1/2$, $c_\phi=1$, see the definitions in Eqs.~(\ref{S}) and~(\ref{Tphi2}). The fermion contribution to $T$ is taken to be $\Delta T_Q=0.1$ in both figures, and the $95\%$ CL limits when $\Delta T_Q=0.2$ are shown for comparison (dot-dashed curves). More details can be found in the text.
\label{EWfit}}
\end{center}
\end{figure}

The large $\Delta S_{\rm UV}>0$ together with the negative $\Delta T_h,\Delta T_\phi$ imply that a good fit can only be achieved if an additional {\emph{positive}} contribution to $T$ is present. This should arise from loops of the top partners in our model, so we take
\ba
\Delta S_{\rm tot}&\equiv&\Delta S_{\rm UV}\\\no
\Delta T_{\rm tot}&\equiv&\Delta T_{h}+\Delta T_\phi+\Delta T_{Q}.
\ea
While $\Delta T_Q$ was not explicitly calculated in this paper, one expects from the discussion in the previous subsections that a value of order $\Delta T_Q\sim+0.1$ be perfectly realistic. We thus present the $68\%, 95\%$, and $99\%$ CL contours of $S,T$ ($U=0$) in figure~\ref{EWfit} assuming $\Delta T_Q=+0.1$. The dot-dashed lines illustrate the $95\%$ CL limit in the case $\Delta T_Q=0.2$. In the figure we took $c_S=1/2$ in~(\ref{S}) and $c_\phi=1$ in~(\ref{Tphi2}), and varied $\xi$ between $\xi=0.2$ (left) and $\xi=0.3$ (right). The best fit values and their correlation are taken from~\cite{Baak:2012kk}.

We did not include in our fit the new effects induced on $\delta g_{b_L}$, because they are expected to be small. Our analysis can therefore provide only a qualitative picture of what a realistic theory would give.

We see in figure~\ref{EWfit} that the vev of the triplet becomes completely negligible as soon as $m_{\phi_+}$ is above a few hundred GeV. Once this threshold is passed the contour is controlled by $\Delta S_{\rm UV}$, and especially $\Delta T_h+\Delta T_{Q}$. We find large regions of the parameter space $m_\rho\gtrsim3.5$ TeV are allowed even with our less optimistic choice of $\Delta T_Q$.

\begin{figure}
\begin{center}
\includegraphics[width=3in]{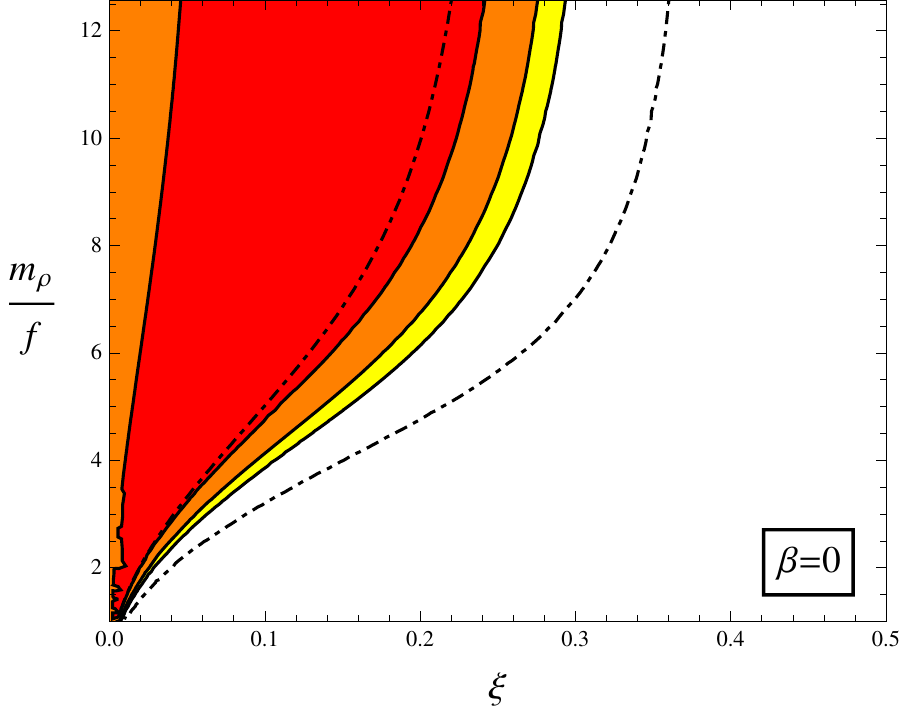}
\caption{\small $68,95,99\%$ CL fit of the parameters $S,T$ ($U=0$) as a function of the coupling $m_\rho/f\lesssim4\pi$ and $\xi$ in the conservative limit $m_{\phi_+}^2=8\gamma/f^2=2m_h^2(2-\xi)/\xi$ (i.e $\beta=0$). In the plot we took $c_S=1/2$, $c_\phi=1$, see the definitions in Eqs.~(\ref{S}) and~(\ref{Tphi2}). The fermion contribution to $T$ is taken to be $\Delta T_Q=0.1$, and the $95\%$ CL limits when $\Delta T_Q=0.2$ is shown by the dot-dashed curve. 
\label{EWfit2}}
\end{center}
\end{figure}

The biggest contribution to the mass of $\phi_+$ comes from $\gamma$, which is the same coupling controlling the smallness of $\xi$. In the limit $\beta=0$ one has $m_{\phi_+}^2=8\gamma/f^2$, and the mass of $\phi_+$ may be written as a function of $\xi,m_h$. We plot the resulting EW fit in figure~\ref{EWfit2} as a function of $\xi$ and $m_\rho/f\lesssim4\pi$ for $\Delta T_Q=0.1$ (as usual the $95\%$ CL bound for $\Delta T_Q=0.2$ is in dot-dashed). Note that the value $\xi=0$ gets disfavored as $\Delta T_Q$ moves to larger positive values because one needs a positive contribution to $S$ to lie in the $\chi^2$ ellipse, and according to~(\ref{S}) this would be absent in the $\xi=0$ limit.

On the one hand, the qualitative physical picture emerging from figure~\ref{EWfit2} is robust. Indeed, EW precision data want to decouple $\phi_+$ and simultaneously have $a^2$ as close to $1$ as possible, and therefore have the tendency to push $f$ to larger scales than direct LHC bounds on the Higgs couplings. On the other hand, the bounds shown in the figure are quantitatively conservative since they ignore other positive corrections to $m_{\phi_+}$. They are also not as robust as the bound of figure~\ref{FitXi} because they heavily rely on the model-dependent parameter $\Delta T_Q$.

Despite the presence of a negative tree-level correction to $T$ from the vev of $\phi_+$, for a given $f$ and $\Delta T_Q$ the EW fit in the $SU(5)/SO(5)$ model is still slightly better compared to that of the minimal cosets $SO(5)/SO(4)$, $SU(4)/Sp(4)$. The point is that minimal scenarios have no triplet problem but have a $\Delta T_h$ four times larger. Numerically, we find that the negative contribution $\Delta T_\phi$ is partially compensated by the gain in $\Delta T_h$. This can be seen by noting that the $SO(5)/SO(4)$ analog of figure~\ref{EWfit2} has the filled area shifted below $\xi=0.2$.

\section{Conclusion}
\label{sec:conclusions}

This paper was motivated by a very simple question:

``Given the increasingly stringent bounds on the top partners as well as the Higgs boson properties, can pseudo-NGB Higgs scenarios still offer a fully natural description of electroweak symmetry breaking?"

To directly address this question we focused on a model based on the coset $SU(5)/SO(5)$. Besides the associated NGBs, the only new fields appearing in our EFT are the {\emph{light, weakly coupled}} fermionic parterns of the top, taken to transform under the fundamental representation of $SU(5)$. We did not include the {\emph{weakly coupled}} gauge boson partners since these fields are not of immediate relevance for the current collider searches and precision measurements. 

The present study suggests the answer to the above question is affirmative: the pseudo-NGB Higgs can still accomplish the purpose it was designed to address. This is pictorially illustrated in figure~\ref{naturalness}, where we show the allowed parameter space after a conservative, flat bound on the top partners from LHC direct searches is imposed. Virtually no tuning is needed to simultaneously explain the Higgs mass and the weak scale. While this specific model is certainly not free from defects, it has the merit of showing that the Natural pseudo-NGB Higgs framework still represents a compelling alternative to Natural SUSY.

The main problem of the coset $SU(5)/SO(5)$ seems to be the presence of a vacuum expectation value for the electroweak triplet NGB $\phi_+$. Yet, we have shown this issue can efficiently be solved by the very same physics responsible for generating a parametrically large Higgs quartic coupling.

This one issue should be weighed along with a number of remarkable features that characterize the symmetry breaking pattern $SU(5)\to SO(5)$:

\begin{itemize}
\item Given a new physics scale $f$, the couplings of the physical Higgs boson to the vector bosons and fermions of the SM are closer to those of a fundamental Higgs than in models based on minimal cosets. Interestingly, $SU(5)/SO(5)$ can be in good agreement with the latest LHC data as soon as $f\gtrsim350$  GeV, see figure~\ref{FitXi}. The importance of this fact in the context of indirect precision measurements has also been emphasized in section~\ref{sec:fit}.
\item Collective breaking of the Nambu-Goldstone symmetry is maximally efficient. This implies that we can push the masses of the top partners above the current LHC bounds ($m_Q\gtrsim700-800$ GeV) compatibly with the naturalness criteria, see figure~\ref{naturalness}.
\item A Higgs quartic coupling is automatically generated via UV-sensitive loops of EW doublet fermions mixing with the top partners, including the SM quark doublet $q$ (section~\ref{sec:quartic}) and possible exotic states (see section~\ref{sec:triplet} and Appendix~\ref{app:quartic}). This result heavily uses the presence of additional NGBs (i.e. of a non-minimal coset), and is needed to explain $v<f$ naturally.
\item The collider signature of the top partners depends crucially on the decays of the exotic NGBs (see section~\ref{sec:collider}). This is typically characterized by a larger jet activity than in the more minimal frameworks, and in generic regions of the parameter space even by a suppression of final states with hard leptons. While not true in our most minimal scenario, it is important to realize that the phenomenology of the main players in the naturalness game could qualitatively depart from the standard one.
\end{itemize}

An important piece of our work is the systematic analysis of the low energy Lagrangian for the SM and the top partners $Q$, in which {\emph{both}} the SM fermions and the $Q$s appear as Partially Composite states of the strong dynamics (section~\ref{sec:picture}). Our scenario has two more expansion parameters compared to the standard Partial Compositeness framework: one controls all the couplings between the Higgs sector and the $Q$s, and the other the ``direct" couplings between the strong dynamics and the SM fermions. We argued this latter must be very small in order to realize an efficient collective breaking of the NGB symmetry. The former should be non-zero in order to generate the SM fermion masses, but small enough to suppress the masses of the top partners compared to the compositeness scale and avoid large flavor-violating effects originating after the mixing with the SM is taken into account. At leading order the resulting picture typically reduces to a ``Little Higgs" scenario, but important corrections are found in the scalar potential.


\acknowledgments
It is a pleasure to acknowledge fruitful discussions with Kaustubh Agashe, Brando Bellazzini, Roberto Franceschini, Yevgeny Kats, Giuliano Panico, Javi Serra, Ian Shoemaker, Daniel Stolarski, and Raman Sundrum. This work was supported in part by the NSF Grant No. PHY-0968854, No. PHY-0910467, and by the Maryland Center for Fundamental Physics.

\appendix

\section{(Truncated) EFT below $m_Q$}

Integrating out the top partners at tree-level from~(\ref{EFT1}) we obtain
\ba\label{LET}
{\cal L}_{\rm EFT}^{\mu<m_Q}&=&\frac{f^2}{4}{\rm tr}[D_\mu U D^\mu U^\dagger]-\delta V_{\rm EFT}\\\no
&+&\overline{q} \gamma^\mu\left[1+\lambda_q (m_Q^\dagger m_Q)^{-1} \lambda_q^\dagger\right] iD_\mu q\\\no
&+&\overline{u} \gamma^\mu\left[1+{\lambda_u} (m_Q m_Q^\dagger)^{-1}{\lambda_u}^\dagger\right] iD_\mu u\\\no
&-&\lambda_q\overline{q}m_Q^{-1}U^\dagger u\lambda_u^\dagger+{\rm hc}\\\no
&+&\overline{q} \gamma^\mu \left[\lambda_q (m_Q^\dagger m_Q)^{-1} (U^\dagger iD_\mu U)\lambda_q^\dagger\right] q\\\no
&+&\overline{u} \gamma^\mu \left[\lambda_u (m_Q m_Q^\dagger)^{-1} (U iD_\mu U^\dagger)\lambda_u^\dagger\right] u\\\no
&+&O(p^2\overline{f_{\rm SM}}f_{\rm SM}),
\ea
Here $(m_Q^\dagger m_Q)^{-1}$ and $(m_Q m_Q^\dagger)^{-1}$ are matrices in flavor space.

\section{Mass Matrix $M_{2/3}$}
\label{app:Mass}

Limiting our analysis to the third generation we find that the mass matrix~(\ref{M23}) for the charge $2/3$ fermions satisfies
\ba\label{det}
{\rm det}(m^2-M^\dagger_{2/3}M_{2/3})&=&(m^2-m_Q^2)^2(m^2-m_Q^2-\lambda_q^2)\\\no
&\times&\left[m^2(m^2-m_Q^2-\lambda_q^2)(m^2-m_Q^2-\lambda_u^2)-\frac{\sin^2}{2}\lambda_q^2\lambda_u^2\right].
\ea
As expected, when $\sin=0$ there is a massless top, a doublet of mass $m_{Q_+}=m_Q$, a doublet of mass $m_{Q_-}=\sqrt{m_Q^2+\lambda_q^2}$, and a singlet of mass $m_{Q_0}=\sqrt{m_Q^2+\lambda_u^2}$. The eigenvalues for $\sin\neq0$ are easily derived in an expansion in $\sin^2$:
\ba
m_{Q_+^u}^2&=&m_Q^2\\\no
m_{Q_+^d}^2&=&m_Q^2\\\no
m_{Q_-^u}^2&=&m_Q^2+\lambda_q^2+\frac{\sin^2}{2}m_Q^2\frac{\lambda_u^2\lambda_q^2}{(m_Q^2+\lambda_q^2)(\lambda_q^2-\lambda_u^2)}+O(\sin^4)\\\no
m_{Q_-^d}^2&=&m_Q^2+\lambda_q^2\\\no
m_{Q_0}^2&=&m_Q^2+\lambda_u^2+\frac{\sin^2}{2}m_Q^2\frac{\lambda_u^2\lambda_q^2}{(m_Q^2+\lambda_u^2)(\lambda_u^2-\lambda_q^2)}+O(\sin^4)\\\no
m_t^2&=&\frac{\sin^2}{2}m_Q^2\frac{\lambda_u^2\lambda_q^2}{(m_Q^2+\lambda_q^2)(m_Q^2+\lambda_u^2)}+O(\sin^4).
\ea
Note that the expansion for the heavy fermions break down when $\lambda_q\sim\lambda_u$, however the potential~(\ref{toppotential}) is smooth in this limit. Confronting with the exact solution we find that the above expression for $m_t$ is always accurate to a level better than $8\%$. As a result, figure~\ref{rates} is accurate as well.

From~(\ref{det}), and making use of the Higgs low energy theorems~\cite{Ellis:1975ap}, we can see that the correction to the $ggh$ vertex from the $Q$s is:
\ba\label{hgg}
v\frac{\partial}{\partial h}\log{\rm det}M_{2/3}=v\frac{\partial \log \sin}{\partial h}=\frac{g_{hff}}{g_{hff}^{\rm SM}}.
\ea
The corrections induced by integrating out the triplet~\cite{Low:2009di}\cite{Low:2010mr} are proportional to $\alpha_{\rm em}\Delta T_\phi/\xi$ and will be small in a realistic (and natural) model.

\section{A quartic from loops of Exotic Fermions}
\label{app:quartic}

A $A_1$-invariant potential for the Higgs may be obtained by imposing $T$-parity~\cite{Cheng:2004yc}(see also~\cite{Low:2004xc}\cite{Hubisz:2004ft}\cite{Pappadopulo:2010jx}). Here we decide to follow a similar, but more minimal approach. This is suggested by our discussion of the Higgs potential in section~\ref{sec:quartic}, and relies on the introduction of exotic fermions in incomplete $SU(5)$ representations.

Suppose the strong dynamics respects an unbroken factorized $SO(2)$~\footnote{This symmetry cannot be identified with $U(1)_X$ because it will be explicitly broken, see below.} and introduce a pair of Dirac fermions $\Psi^t=(\psi_1,\psi_2)$ transforming as a $(5,2)$ of $SU(5)\times SO(2)$. The two sectors couple again via the $Q$s:
\ba
\lambda_\Psi\overline{\Psi}Q_L + m_\Psi\overline{\Psi}\Psi,
\ea
where $m_\Psi$ is a universal mass and we neglected a possible coupling for $Q_R$ for simplicity. 

As it stands, the model does not break any of the symmetries, and in particular gives no contribution to the Higgs potential. We thus add a soft breaking term:
\ba
\delta m_\Psi\overline{\Psi}\Psi
\ea
which gives a mass to all components except $\psi_1^-,\psi_2^+$. In the regime $m_\Psi\ll\delta m_\Psi$ that we will be interested in, our choice may be made radiatively stable by invoking an approximate chiral symmetry for the light components. 

A non-vanishing $\delta m_\Psi$ breaks both $SO(2)$ and $SU(5)$. At energy scales much below $\delta m_\Psi$ we integrate out the heavy fermions and observe that the symmetry breaking effects are entirely encoded in $\lambda_\Psi$ and $m_\Psi$, up to corrections suppressed by powers of $\lambda_\Psi/\delta m_\Psi$. This can be made explicit by keeping track of the spurionic symmetries. We find it convenient to write 
\ba
\delta m_\Psi=(1_{10\times10}-\Delta\Delta^\dagger)(M_\Psi-m_\Psi),~~~~~~~~~~~~~\Delta\equiv\left( \begin{array}{cc}  \Delta_-^\dagger\Delta_- &~ \\
~&\Delta_+^\dagger\Delta_+\end{array}\right),
\ea
where $M_\Psi\gg m_\Psi$ has an unspecified flavor structure, and $\Delta_{\pm}$ have the following numerical values
\ba
\Delta_-=\left({\bf 1}_{2\times2}\quad{\bf 0}_{2\times2}\quad{\bf 0}_{2\times1}\right)~~~~~~~~~~~~~~\Delta_+=\left({\bf 0}_{2\times2}\quad{\bf 1}_{2\times2}\quad{\bf 0}_{2\times1}\right).
\ea
In terms of $\Delta$ the total $\Psi$ mass term acquires the useful expression $m_\Psi+\delta m_\Psi=m_\Psi \Delta\Delta^\dagger+(1-\Delta\Delta^\dagger)M_\Psi$. Now the effective field theory at energies $\ll M_\Psi$ can be easily written down
\ba
\lambda_\Psi\overline{\Psi}\Delta Q_L + m_\Psi  \overline{\Psi}\Delta\Delta^\dagger{\Psi}+{\rm powers~of}~\frac{\lambda}{M_\Psi}.
\ea
We will view $\Delta$ as a background field transforming as $\Delta\to U_\Psi\Delta U^\dagger_Q$ under the spurionic symmetries $U_\alpha\in SU(5)_\alpha\times SO(2)_\alpha$ (with $\alpha=\Psi,Q$). That large symmetry was originally broken down to its diagonal subgroup by $\lambda_\Psi$.

As expected, we see that $\Delta_-\neq\Delta_+$ implies $SO(2)$ breaking, while $\Delta^\dagger_\pm\Delta_\pm\neq1_{5\times5}$ tells us that $SU(5)$ is broken. Loops of the light $\psi_1^-,\psi_2^+$ will therefore contribute to the scalar potential. Because this effect decouples as powers of $m_\Psi\to\infty$ we will assume $m_\Psi\lesssim m_\rho$. In this case, one loop diagrams will be typically dominated by scales of order $m_\rho$ and the potential will be a trace of some function of the following invariant 
\ba\label{solution}
\Delta^\dagger\Delta U\Sigma_0\left(\Delta^\dagger\Delta U\Sigma_0\right)^*.
\ea
The structure in~(\ref{solution}) is precisely the same as in~(\ref{structure'}). The key observation is that $A_1$ acts as $\Delta_-^\dagger\Delta_-\leftrightarrow\Delta_+^\dagger\Delta_+$ and can now be compensated by $SO(2)$ rotations, so~(\ref{solution}) is $A_1$-invariant. In complete analogy with~(\ref{quartic}), the potential generated from it renormalizes the quartic Higgs coupling $\gamma$ and $m_{\phi_+}$, but not the Higgs mass.

In models where~(\ref{solution}) provides the dominant contribution to $\gamma$ the electroweak $T$ parameter reads as in~(\ref{Tphi2}). There is also a 1-loop contribution from the exotic fermions, but this is well under control if $m_\Psi$ is large enough compared to $v$.

Before concluding we would like to make a couple of comments. First of all, the component $\psi_2^+$ mixes with $Q_+$, potentially affecting its collider signatures, and with it the entire phenomenology of the model (see section~\ref{sec:collider}). Second, it is possible to consider more minimal scenarios in which $\psi_1^-$ is identified with the SM quark doublet $q$, though we will not attempt to do it here.



\begin{thebibliography}{99}
 
 
\bibitem{Higgs!} 
  G.~Aad {\it et al.}  [ATLAS Collaboration],
  Phys.\ Lett.\ B {\bf 716}, 1 (2012)
  [arXiv:1207.7214 [hep-ex]].
  
  
  S.~Chatrchyan {\it et al.}  [CMS Collaboration],
  Phys.\ Lett.\ B {\bf 716}, 30 (2012)
  [arXiv:1207.7235 [hep-ex]].
  
  
  
  \bibitem{HiggsFits}
  
    
  Here is an incomplete list of recent papers on the subject.
  
    

  
  
  
  D.~Carmi, A.~Falkowski, E.~Kuflik, T.~Volansky and ,
  JHEP {\bf 1207}, 136 (2012)
  [arXiv:1202.3144 [hep-ph]].
  
  
  
  
  
  
  
  
  
  
  
  
  
  A.~Azatov, R.~Contino, J.~Galloway and ,
  JHEP {\bf 1204}, 127 (2012)
  [arXiv:1202.3415 [hep-ph]].
  
  
  
  
  
  
  
  
  J.~R.~Espinosa, C.~Grojean, M.~Muhlleitner, M.~Trott and ,
  JHEP {\bf 1205}, 097 (2012)
  [arXiv:1202.3697 [hep-ph]].
 
  
  M.~Klute, R.~Lafaye, T.~Plehn, M.~Rauch, D.~Zerwas and ,
  Phys.\ Rev.\ Lett.\  {\bf 109}, 101801 (2012)
  [arXiv:1205.2699 [hep-ph]].
  
  
  P.~P.~Giardino, K.~Kannike, M.~Raidal and A.~Strumia,
  Phys.\ Lett.\ B {\bf 718}, 469 (2012)
  [arXiv:1207.1347 [hep-ph]].
  
  
  
  J.~Ellis and T.~You,
  JHEP {\bf 1209}, 123 (2012)
  [arXiv:1207.1693 [hep-ph]].

  
    
  J.~R.~Espinosa, C.~Grojean, M.~Muhlleitner and M.~Trott,
  JHEP {\bf 1212}, 045 (2012)
  [arXiv:1207.1717 [hep-ph]].
  
      
  T.~Plehn, M.~Rauch and ,
  Europhys.\ Lett.\  {\bf 100}, 11002 (2012)
  [arXiv:1207.6108 [hep-ph]].
  
     
  T.~Corbett, O.~J.~P.~Eboli, J.~Gonzalez-Fraile, M.~C.~Gonzalez-Garcia and ,
  Phys.\ Rev.\ D {\bf 87}, 015022 (2013)
  [arXiv:1211.4580 [hep-ph]].


  J.~Reuter and M.~Tonini,
  JHEP {\bf 1302}, 077 (2013)
  [arXiv:1212.5930 [hep-ph]].
  
  X.~-F.~Han, L.~Wang, J.~M.~Yang and J.~Zhu,
  arXiv:1301.0090 [hep-ph].


  A.~Falkowski, F.~Riva, A.~Urbano and ,
  arXiv:1303.1812 [hep-ph].
  
  T.~Alanne, S.~Di Chiara and K.~Tuominen,
  arXiv:1303.3615 [hep-ph].
  
  
   A.~Djouadi, G.~Moreau and ,
  arXiv:1303.6591 [hep-ph].




\bibitem{Montull:2012ik} 
  M.~Montull and F.~Riva,
  JHEP {\bf 1211}, 018 (2012)
  [arXiv:1207.1716 [hep-ph]].

  
  
  
\bibitem{Giardino:2013bma} 
  P.~P.~Giardino, K.~Kannike, I.~Masina, M.~Raidal and A.~Strumia,
  arXiv:1303.3570 [hep-ph].
  



  
  
  
  
  
  
  
  
  
  
  
  
  
  
  
  
  
  
  
  
  
 
 
 \bibitem{naturalness} 
 G. Õt Hooft, Naturalness, chiral symmetry, and spontaneous chiral symmetry breaking. Lecture given at Cargese Summer Institute, Cargese, France, Aug 26 - Sep 8, 1979.
 
 
\bibitem{Barbieri:1987fn} 
  R.~Barbieri, G.~F.~Giudice and ,
  Nucl.\ Phys.\ B {\bf 306}, 63 (1988).
 
 
\bibitem{CH} 
  D.~B.~Kaplan, H.~Georgi and ,
  Phys.\ Lett.\ B {\bf 136}, 183 (1984).
  
  D.~B.~Kaplan, H.~Georgi, S.~Dimopoulos and ,
  Phys.\ Lett.\ B {\bf 136}, 187 (1984).
  
  
  
      
  
\bibitem{Berger:2012ec} 
  J.~Berger, J.~Hubisz, M.~Perelstein and ,
  JHEP {\bf 1207}, 016 (2012)
  [arXiv:1205.0013 [hep-ph]].
  
  
  
    
  
\bibitem{Matsedonskyi:2012ym} 
  O.~Matsedonskyi, G.~Panico, A.~Wulzer and ,
  JHEP {\bf 1301}, 164 (2013)
  [arXiv:1204.6333 [hep-ph]].
  
  
\bibitem{Panico:2012uw} 
  G.~Panico, M.~Redi, A.~Tesi, A.~Wulzer and ,
  JHEP {\bf 1303}, 051 (2013)
  [arXiv:1210.7114 [hep-ph]].


\bibitem{DeSimone:2012fs} 
  A.~De Simone, O.~Matsedonskyi, R.~Rattazzi, A.~Wulzer and ,
  arXiv:1211.5663 [hep-ph].
  
  
\bibitem{Kaplan:1991dc} 
  D.~B.~Kaplan,
  Nucl.\ Phys.\ B {\bf 365}, 259 (1991).
  
  
  
\bibitem{Gherghetta:2000qt} 
  T.~Gherghetta, A.~Pomarol and ,
  Nucl.\ Phys.\ B {\bf 586}, 141 (2000)
  [hep-ph/0003129].
  
  
  
\bibitem{Huber:2000ie} 
  S.~J.~Huber, Q.~Shafi and ,
  Phys.\ Lett.\ B {\bf 498}, 256 (2001)
  [hep-ph/0010195].
  
  
  
\bibitem{KerenZur:2012fr} 
  B.~Keren-Zur, P.~Lodone, M.~Nardecchia, D.~Pappadopulo, R.~Rattazzi and L.~Vecchi,
  Nucl.\ Phys.\ B {\bf 867}, 429 (2013)
  [arXiv:1205.5803 [hep-ph]].
  
\bibitem{Redi:2011zi} 
  M.~Redi and A.~Weiler,
  JHEP {\bf 1111}, 108 (2011)
  [arXiv:1106.6357 [hep-ph]].



  
\bibitem{Barbieri:2012tu} 
  R.~Barbieri, D.~Buttazzo, F.~Sala, D.~M.~Straub and A.~Tesi,
  arXiv:1211.5085 [hep-ph].
  
  
\bibitem{Vecchi:2012fv} 
  L.~Vecchi,
  arXiv:1206.4701 [hep-ph].
  
  
\bibitem{ArkaniHamed:2001nc} 
  N.~Arkani-Hamed, A.~G.~Cohen, H.~Georgi and ,
  Phys.\ Lett.\ B {\bf 513}, 232 (2001)
  [hep-ph/0105239].
  
  
\bibitem{ArkaniHamed:2002qy} 
  N.~Arkani-Hamed, A.~G.~Cohen, E.~Katz, A.~E.~Nelson and ,
  JHEP {\bf 0207}, 034 (2002)
  [hep-ph/0206021].
 

\bibitem{Katz:2003sn} 
  E.~Katz, J.~-y.~Lee, A.~E.~Nelson and D.~G.~E.~Walker,
  JHEP {\bf 0510}, 088 (2005)
  [hep-ph/0312287].

  
\bibitem{Cacciapaglia:2007fw} 
  G.~Cacciapaglia, C.~Csaki, J.~Galloway, G.~Marandella, J.~Terning and A.~Weiler,
  JHEP {\bf 0804}, 006 (2008)
  [arXiv:0709.1714 [hep-ph]].
  
  
\bibitem{Schmaltz:2010ac} 
  M.~Schmaltz, D.~Stolarski and J.~Thaler,
  JHEP {\bf 1009}, 018 (2010)
  [arXiv:1006.1356 [hep-ph]].
  
  
\bibitem{Cheng:2004yc} 
  H.~-C.~Cheng, I.~Low and ,
  JHEP {\bf 0408}, 061 (2004)
  [hep-ph/0405243].
  


  
  
  
\bibitem{Birkedal:2004xi} 
  A.~Birkedal, Z.~Chacko, M.~K.~Gaillard and ,
  JHEP {\bf 0410}, 036 (2004)
  [hep-ph/0404197].
  
  
\bibitem{Chankowski:2004mq} 
  P.~H.~Chankowski, A.~Falkowski, S.~Pokorski, J.~Wagner and ,
  Phys.\ Lett.\ B {\bf 598}, 252 (2004)
  [hep-ph/0407242].
  
  
\bibitem{Roy:2005hg} 
  T.~S.~Roy, M.~Schmaltz and ,
  JHEP {\bf 0601}, 149 (2006)
  [hep-ph/0509357].
  
  
\bibitem{Csaki:2005fc} 
  C.~Csaki, G.~Marandella, Y.~Shirman, A.~Strumia and ,
  Phys.\ Rev.\ D {\bf 73}, 035006 (2006)
  [hep-ph/0510294].
  
  
  
 
  
  
  \bibitem{CMS} 
  [CMS Collaboration] CMS-PAS-B2G-12-012
  
  
  \bibitem{ATLAS} 
  [ATLAS Collaboration] ATLAS-CONF-2013-018

  
  
  
  
\bibitem{Giudice:2007fh} 
  G.~F.~Giudice, C.~Grojean, A.~Pomarol, R.~Rattazzi and ,
  JHEP {\bf 0706}, 045 (2007)
  [hep-ph/0703164].
  

  
  
    
  
\bibitem{Contino:2006qr} 
  R.~Contino, L.~Da Rold and A.~Pomarol,
  Phys.\ Rev.\ D {\bf 75}, 055014 (2007)
  [hep-ph/0612048].
  

  
  
  
\bibitem{Georgi:1984af} 
  H.~Georgi, D.~B.~Kaplan and ,
  Phys.\ Lett.\ B {\bf 145}, 216 (1984).
  
  

  
  
   

\bibitem{Katz:2005au} 
  E.~Katz, A.~E.~Nelson and D.~G.~E.~Walker,
  JHEP {\bf 0508}, 074 (2005)
  [hep-ph/0504252].
  
\bibitem{CCWZ} 
  S.~R.~Coleman, J.~Wess, B.~Zumino and ,
  Phys.\ Rev.\  {\bf 177}, 2239 (1969).
  
  C.~G.~Callan, Jr., S.~R.~Coleman, J.~Wess, B.~Zumino and ,
  Phys.\ Rev.\  {\bf 177}, 2247 (1969).
  
  


\bibitem{Gripaios:2009pe} 
  B.~Gripaios, A.~Pomarol, F.~Riva, J.~Serra and ,
  JHEP {\bf 0904}, 070 (2009)
  [arXiv:0902.1483 [hep-ph]].


  
  
\bibitem{Agashe:2004rs} 
  K.~Agashe, R.~Contino and A.~Pomarol,
  Nucl.\ Phys.\ B {\bf 719}, 165 (2005)
  [hep-ph/0412089].
  

  
  
\bibitem{Galloway:2010bp} 
  J.~Galloway, J.~A.~Evans, M.~A.~Luty, R.~A.~Tacchi and ,
  JHEP {\bf 1010}, 086 (2010)
  [arXiv:1001.1361 [hep-ph]].
  
  
\bibitem{Low:2009di} 
  I.~Low, R.~Rattazzi, A.~Vichi and ,
  JHEP {\bf 1004}, 126 (2010)
  [arXiv:0907.5413 [hep-ph]].
  

  
   
\bibitem{Ellis:1975ap} 
  J.~R.~Ellis, M.~K.~Gaillard, D.~V.~Nanopoulos and ,
  Nucl.\ Phys.\ B {\bf 106}, 292 (1976).




  
\bibitem{Low:2010mr} 
  I.~Low, A.~Vichi and ,
  Phys.\ Rev.\ D {\bf 84}, 045019 (2011)
  [arXiv:1010.2753 [hep-ph]].
  
\bibitem{Azatov:2011qy} 
  A.~Azatov, J.~Galloway and ,
  Phys.\ Rev.\ D {\bf 85}, 055013 (2012)
  [arXiv:1110.5646 [hep-ph]].
  
\bibitem{Gillioz:2012se} 
  M.~Gillioz, R.~Grober, C.~Grojean, M.~Muhlleitner, E.~Salvioni and ,
  JHEP {\bf 1210}, 004 (2012)
  [arXiv:1206.7120 [hep-ph]].
  
\bibitem{Delaunay:2013iia} 
  CŽd.~Delaunay, C.~Grojean, G.~Perez and ,
  arXiv:1303.5701 [hep-ph].
  

\bibitem{Chala:2012af} 
  M.~Chala,
  JHEP {\bf 1301}, 122 (2013)
  [arXiv:1210.6208 [hep-ph]].


\bibitem{Carena:2012xa} 
  M.~Carena, I.~Low, C.~E.~M.~Wagner and ,
  JHEP {\bf 1208}, 060 (2012)
  [arXiv:1206.1082 [hep-ph]].
  

\bibitem{Marzocca:2012zn} 
  D.~Marzocca, M.~Serone, J.~Shu and ,
  JHEP {\bf 1208}, 013 (2012)
  [arXiv:1205.0770 [hep-ph]].
  
  
\bibitem{Pomarol:2012qf} 
  A.~Pomarol, F.~Riva and ,
  JHEP {\bf 1208}, 135 (2012)
  [arXiv:1205.6434 [hep-ph]].
  

 

\bibitem{Han:2003wu} 
  T.~Han, H.~E.~Logan, B.~McElrath, L.~-T.~Wang and ,
  Phys.\ Rev.\ D {\bf 67}, 095004 (2003)
  [hep-ph/0301040].
  
  
\bibitem{Csaki:2003si} 
  C.~Csaki, J.~Hubisz, G.~D.~Kribs, P.~Meade, J.~Terning and ,
  Phys.\ Rev.\ D {\bf 68}, 035009 (2003)
  [hep-ph/0303236].
  
  
\bibitem{Contino:2008hi} 
  R.~Contino, G.~Servant and ,
  JHEP {\bf 0806}, 026 (2008)
  [arXiv:0801.1679 [hep-ph]].
  

  
  
\bibitem{Mrazek:2009yu} 
  J.~Mrazek, A.~Wulzer and ,
  Phys.\ Rev.\ D {\bf 81}, 075006 (2010)
  [arXiv:0909.3977 [hep-ph]].
  


    
  


  

\bibitem{Contino:2006nn} 
  R.~Contino, T.~Kramer, M.~Son and R.~Sundrum,
  JHEP {\bf 0705}, 074 (2007)
  [hep-ph/0612180].
  
  
  
\bibitem{Barbieri:2007bh} 
  R.~Barbieri, B.~Bellazzini, V.~S.~Rychkov, A.~Varagnolo and ,
  Phys.\ Rev.\ D {\bf 76}, 115008 (2007)
  [arXiv:0706.0432 [hep-ph]].
  
  
\bibitem{Anastasiou:2009rv} 
  C.~Anastasiou, E.~Furlan and J.~Santiago,
  Phys.\ Rev.\ D {\bf 79}, 075003 (2009)
  [arXiv:0901.2117 [hep-ph]].
  
\bibitem{Gillioz:2008hs} 
  M.~Gillioz,
  Phys.\ Rev.\ D {\bf 80}, 055003 (2009)
  [arXiv:0806.3450 [hep-ph]].



\bibitem{Agashe:2003zs} 
  K.~Agashe, A.~Delgado, M.~J.~May, R.~Sundrum and ,
  JHEP {\bf 0308}, 050 (2003)
  [hep-ph/0308036].



  
  
\bibitem{Agashe:2006at} 
  K.~Agashe, R.~Contino, L.~Da Rold and A.~Pomarol,
  Phys.\ Lett.\ B {\bf 641}, 62 (2006)
  [hep-ph/0605341].
  
  
\bibitem{Mrazek:2011iu} 
  J.~Mrazek, A.~Pomarol, R.~Rattazzi, M.~Redi, J.~Serra, A.~Wulzer and ,
  Nucl.\ Phys.\ B {\bf 853}, 1 (2011)
  [arXiv:1105.5403 [hep-ph]].
  
  
 
  
\bibitem{Panico:2011pw} 
  G.~Panico and A.~Wulzer,
  JHEP {\bf 1109}, 135 (2011)
  [arXiv:1106.2719 [hep-ph]].
  

\bibitem{Baak:2012kk} 
  M.~Baak, M.~Goebel, J.~Haller, A.~Hoecker, D.~Kennedy, R.~Kogler, K.~Moenig and M.~Schott {\it et al.},
  Eur.\ Phys.\ J.\ C {\bf 72}, 2205 (2012)
  [arXiv:1209.2716 [hep-ph]].
    


  

    
\bibitem{Low:2004xc} 
  I.~Low,
  JHEP {\bf 0410}, 067 (2004)
  [hep-ph/0409025].
  
\bibitem{Hubisz:2004ft} 
  J.~Hubisz, P.~Meade and ,
  Phys.\ Rev.\ D {\bf 71}, 035016 (2005)
  [hep-ph/0411264].
  
  

  
    
  
\bibitem{Pappadopulo:2010jx} 
  D.~Pappadopulo, A.~Vichi and ,
  JHEP {\bf 1103}, 072 (2011)
  [arXiv:1007.4807 [hep-ph]].
  


    
    
    


 \end{thebibliography}
 \end{document}